\documentclass[12pt]{article}
\usepackage{graphicx, amsmath, amssymb, cite, setspace, color,relsize,bm,bbold,dsfont}

\usepackage[top=1 in, bottom=1 in, left=.9 in, right=.9 in]{geometry}

\newcommand{\captionfonts}{\footnotesize} 
\makeatletter 
\long\def\@makecaption#1#2{%
  \vskip\abovecaptionskip
  \sbox\@tempboxa{{\captionfonts #1: #2}}%
  \ifdim \wd\@tempboxa >\hsize
    {\captionfonts #1: #2\par}
  \else
    \hbox to\hsize{\hfil\box\@tempboxa\hfil}%
  \fi
  \vskip\belowcaptionskip}
\makeatother

\def\fnote#1#2{\begingroup\def\thefootnote{#1}\footnote{#2}
     \addtocounter{footnote}{-1}\endgroup}

\usepackage{titlesec}

\titleformat{\paragraph}
{\normalfont\normalsize\bfseries}{\theparagraph}{1em}{}
\titlespacing*{\paragraph}
{0pt}{3.25ex plus 1ex minus .2ex}{1.5ex plus .2ex}

\begin{document}

\title{Spectrum and Stability of\\ Compactifications on Product Manifolds}
\date{}

\author{Adam~R.~Brown$^{1}$ and Alex~Dahlen$^{2}$ \vspace{.1 in}\\
 \vspace{-.3 em}  $^1$ \textit{\small{Physics Department, Stanford University, Stanford, CA 94305, USA}} \\
 \vspace{-.3 em}  $^2$ \textit{\small{Berkeley Center for Theoretical Physics, Berkeley, CA 94720, USA}} }
\date{}

\maketitle
\fnote{}{\hspace{-.65cm}emails: \tt{adambro@stanford.edu, adahlen@berkeley.edu}}
\vspace{-.95cm}

\maketitle

\begin{abstract}
\noindent 
We study the spectrum and perturbative stability of Freund-Rubin compactifications on $M_p \times M_{Nq}$, where $M_{Nq}$ is itself a product of $N$ $q$-dimensional Einstein manifolds.  The higher-dimensional action has a cosmological term $\Lambda$ and a $q$-form flux, which individually wraps each element of the product; the extended dimensions $M_p$ can be anti-de Sitter, Minkowski, or de Sitter.  We find the masses of every excitation around this background, as well as the conditions under which these solutions are stable.  This generalizes previous work on Freund-Rubin vacua, which focused on the $N=1$ case,  in which a $q$-form flux wraps a single $q$-dimensional Einstein manifold.  The $N=1$ case can have a classical instability when the $q$-dimensional internal manifold is a product---one of the members of the product wants to shrink while the rest of the manifold expands.  Here, we will see that individually wrapping each element of the product with a lower-form flux cures this cycle-collapse instability.  The $N=1$ case can also have an instability when $\Lambda>0$ and $q\ge4$ to shape-mode perturbations; we find the same instability in compactifications with general $N$, and show that it even extends to cases where $\Lambda\le0$.  On the other hand, when $q=2$ or 3, the shape modes are always stable and there is a broad class of AdS and de Sitter vacua that are perturbatively stable to all fluctuations.
\end{abstract}

\thispagestyle{empty} 
\newpage

\section{Introduction}

Compactifications that rely on flux to buttress the extra dimensions against collapse were first studied by Freund and Rubin \cite{Freund:1980xh}.  The simplest such models invoke a $q$-form flux, which uniformly wraps a $q$-dimensional internal Einstein manifold.  The stability and spectrum of these compactifications were studied in a series of classic papers \cite{Salam:1981xd, vanNieuwenhuizen:1984iz, Kim:1985ez, DeWolfe:2001nz, Bousso:2002fi}.

In this paper, we will look at generalizations of these simple compactifications to the case where the internal manifold is a product of $N$ $q$-dimensional Einstein manifolds, and each sub-manifold is individually wrapped by a $q$-form flux.  We find the spectrum of small fluctuations around these backgrounds as well as the conditions  for stability.    Our motivations are four-fold.


First, the $N=1$ case has a perturbative instability when the internal $q$-dimensional manifold is itself a product, and the $q$-form flux collectively wraps the entire product. The fluctuation mode in which one element of the product shrinks while the rest of the manifold grows can be unstable.  We will see that moving to higher $N$ explicitly stabilizes that mode; when each element of an internal product manifold is individually wrapped by a lower-form flux, the solution is stable against cycle collapse.

Second, this analysis covers a wide class of interesting models.  Compactifications of string theory down to four dimensions often take the internal manifold to be a product of Einstein manifolds \cite{Witten:1984dg, Witten:1985xc}, so a general study of their spectrum can teach us about low-energy physics.  Furthermore, compactifications with non-trivial internal topology can open up new possibilities in the study of AdS/CFT.  For instance, there is a 27-dimensional bosonic M-theory with a 4-form flux \cite{Horowitz:2000gn} which has compactifications down to AdS$_{27-4N}\times ($S$_4)^N$; the spectrum found in this paper should match to a CFT dual.  The results in this paper would be straightforward to extend to supersymmetric setups as well, allowing for the study of compactifications of IIB string theory such as AdS$_4\times(S_3)^2$, which our results suggest are stable, or compactifications of M-theory  such as AdS$_3\times(S_4)^2$, which our results suggest may be unstable.


Third, these product manifold setups give rise to landscapes that, because of their complexity, serve as interesting  toy models of the string theory landscape while at the same time, because they are made of such simple ingredients, still allow for direct computation.  The number of flux lines wrapping each cycle is quantized and quantum nucleations of charged branes mediate transitions between the vacua.  An upcoming paper will show that such landscapes generically have a large or even unbounded number of de Sitter vacua \cite{Brown:2013fba,Brown:2014sba}.  

Finally, this paper represents an extension of computational methods to a case where the background fields are not uniform over the internal manifold.  If the sub-manifolds differ in the amount of flux that wraps them, then they will also differ in their curvature, so their product will not be an Einstein manifold (will not have $R_{\alpha\beta}$ proportional to $g_{\alpha\beta}$).  Studying these non-uniform compactifications involves developing new techniques; we give an explicit decomposition of the modes into transverse and longitudinal eigenvectors of the Laplacian restricted to each sub-manifold.  Under this decomposition, the modes decouple and we can read off the spectrum.  

\subsection{Review of the $N=1$ Case}
\label{N1}

Before we discuss compactifications on a product of $N$  individually-wrapped Einstein manifolds, it will be helpful first to review the well-studied case of compactifications on a single Einstein manifold.

The simplest Freund-Rubin compactifications are of the form $M_p \times M_q$, where a $q$-form flux wraps a $q$-dimensional positive-curvature Einstein space.  The $p$ extended dimensions form a maximally symmetric manifold, either de Sitter, Minkowksi, or Anti-de Sitter.  These compactifications are solutions to the equations of motion that follow from the action
\begin{equation}
S=\int d^px d^qy \sqrt{-g} \left( R-\frac1{2q!}F_q^2-2\Lambda\right),
\end{equation}
where $\Lambda$ is a higher-dimensional cosmological constant.  The $q$-form flux is taken to uniformly wrap the extra dimensions
\begin{equation}
F_q=c \;\text{vol}_{M_q},
\end{equation}
where $\text{vol}_{M_q}$ is the volume form of $M_q$ and $c$ is the flux density; for compactifications without warping, this  uniform distribution of flux is the only static solution to Maxwell's equation. 

The study of the stability of these solutions to small fluctuations was done by \cite{DeWolfe:2001nz,Bousso:2002fi} and three classes of instabilities were found:
\begin{itemize}
\item{\textbf{Total-Volume Instability}}:  When $\Lambda>0$, there can be an instability for the total volume of the internal manifold to either grow or shrink.  This instability turns on whenever the density of flux lines wrapping the extra dimensions is too small.
\item{\textbf{Lumpiness Instability}}: When $\Lambda>0$ and the internal manifold is a $q$-sphere with $q\ge4$, there can be instability for the internal sphere to become lumpy.  Depending on the flux density, spherical-harmonic perturbations with angular momentum $\ell\ge2$ can be unstable.
\item{\textbf{Cycle-Collapse Instability}}: When $M_q$ is a product manifold, there may be an instability for part of the manifold to grow, while the rest shrinks down to zero volume. For instance, if a $4$-form flux is wrapped around the product of two 2-spheres, then there is an instability in which one of the spheres grows while the other collapses, but with the total volume being preserved.
\end{itemize}
Let's discuss these instabilities and their endpoints in more detail:

The `total-volume instability' can be understood from the perspective of the effective potential.  When the shape of the internal manifold and the number of flux lines $n$ wrapping it are held fixed, you can describe the total-volume modulus as a field living in an effective potential.  When $\Lambda\le0$, the effective potential resembles the left panel of Fig.~1, with a single AdS minimum ($V_\text{eff}<0$).  As the number of flux lines is increased, the minimum moves rightward to larger volumes, and upward to less negatively curved AdS spacetimes.  The flux density $c$ is equal to the number of flux lines $n$ divided by the volume; it turns out that increasing $n$ shifts the minimum far enough to the right that even though the number of flux lines is increased, the density of flux lines around the manifold decreases---increasing $n$ decreases $c$.  When $\Lambda>0$, the effective potential looks qualitatively different, as shown in the right panel of Fig.~1: instead of having one extremum, the potential now has either two or zero extrema.  When the number of flux lines $n$ is small enough, the potential has a minimum at small volume and a maximum at large volume.  Increasing the number of flux lines past a critical value causes the minimum and maximum to merge and annihilate.  Small volume means large flux density, so the large $c$ solutions correspond to minima where the total volume mode is explicitly stabilized and small $c$ solutions correspond to maxima where the total volume mode is explicitly unstable.  The endpoint of this instability is therefore either flow out towards decompactification or flow in towards the stable minimum, where the volume is smaller and the flux density is correspondingly larger \cite{Krishnan:2005su}.

\begin{figure}[t] 
   \centering
   \includegraphics[width=\textwidth]{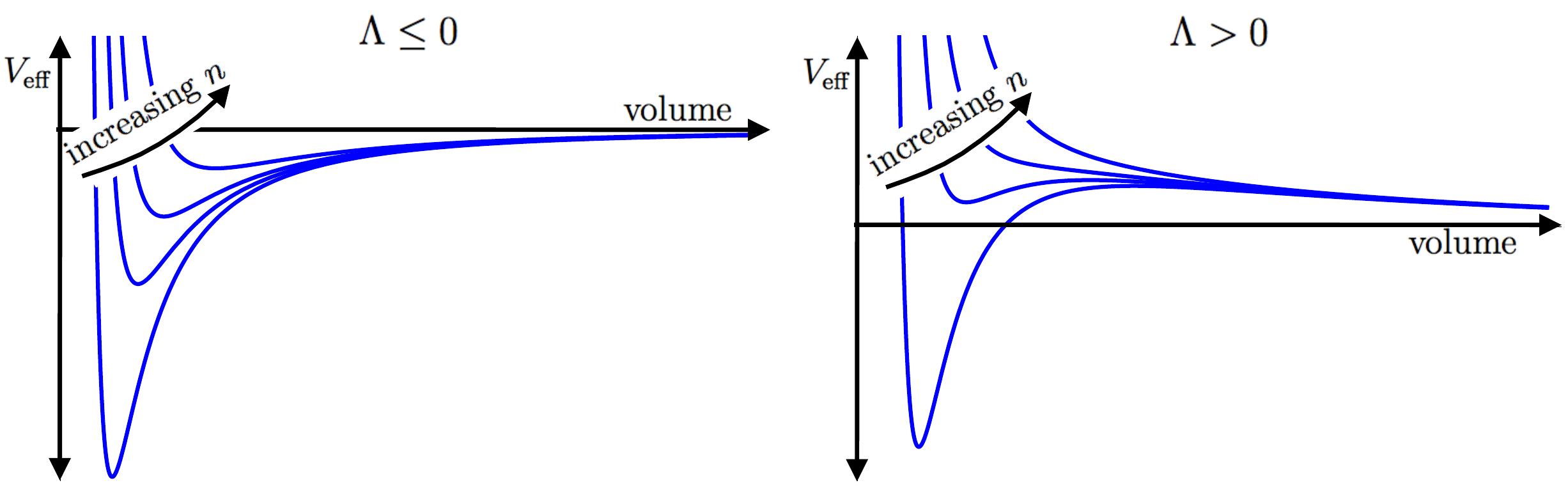} \label{fig:fig1}
   \caption{The Einstein-frame effective potential $V_\text{eff}$ for the total volume behaves differently depending on the sign of the higher-dimensional cosmological constant $\Lambda$.  If $\Lambda\le0$, then the potential has a single AdS minimum for any number of flux lines $n$.  If $\Lambda>0$, then the potential has either two or zero extrema depending on the magnitude of $n$; the maximum is always dS, while the minimum can be either AdS, Minkowksi, or dS.  The `total-volume instability' corresponds to being a maximum and not a minimum on this plot.}
   \label{fig:ManyFluxFlat}
\end{figure}

The `lumpiness instability', like the `total-volume instability', only exists when $\Lambda>0$.  The instability is perhaps surprising because the normal intuition that comes from banging drums and plucking strings is that higher modes have a higher mass and play a higher pitch.  The `lumpiness instability', however, defies this intuition: the total-volume mode, with angular momentum $\ell=0$, can be stable while a mode with $\ell\ge2$ can have a negative mass squared.  The origin of the instability, mathematically, is the coupling of metric modes and flux modes. For the higher-angular-momentum perturbations ($\ell\ge2$), the two modes form a coupled system that needs to be diagonalized, while for the $\ell=0$ and $\ell=1$ modes, only  a single mode enters and no diagonalization is necessary.  (Roughly, the $\ell=0$ mode of the flux potential perturbations is gauge because the total number of flux lines is a conserved quantity; and the $\ell=1$ mode of the metric perturbations is gauge because shifting a sphere by an $\ell=1$ spherical harmonic gives a translated sphere, which has the same induced metric.)  Diagonalization mixes the shape and flux modes and can give rise to negative mass squareds.  When $\Lambda\le0$, these negative mass squareds all lie above the BF bound \cite{Breitenlohner:1982bm}, but when $\Lambda>0$ instabilities can appear; the details are as follows:
\begin{itemize}
\item When $q=2$, all higher-mode fluctuations have a positive mass.
\item When $q=3$, all higher-mode fluctuations have a stable mass squared.
\item When $q\ge4$ and $\Lambda>0$, higher-mode fluctuations can be unstable.
\end{itemize}
 If a solution flows down this instability, where does it land?
Warped compactifications on prolate, oblate, and otherwise lumpy spheres have been found numerically \cite{Kinoshita:2007uk, Kinoshita:2009hh, Lim:2012gh}, analogous to the lumpy black string solutions found  in the study of the Gregory-Laflamme instability \cite{Gubser:2001ac,Gregory:1993vy}; these warped compactifications likely are the endpoint of this flow, but their stability has not been checked, so it is not known whether they are minima or saddle points of the effective potential.

The `cycle-collapse instability', unlike the first two, can exist not only when $\Lambda>0$ but also when $\Lambda=0$ or $\Lambda<0$.  Earlier discussions of this instability appeared in \cite{Duff:1984sv, Berkooz:1998qp, Yasuda:1984py}.  The `cycle-collapse instability' is related to the fact that the compactification uses a highest-form flux to stabilize the internal manifold: a $q$-form flux around a $q$-dimensional manifold.  Highest-form fluxes aren't sensitive to sub-curvatures, so when one sub-manifold gets a little smaller and the other gets a little bigger, the flux can't react to pull things back.  Wrapping flux around the internal manifold has stabilized the total volume, but it has not stabilized the volume of each sub-manifold individually.   
What is the endpoint of this instability?  
The endpoint cannot be that the collapsing cycle stabilizes at a smaller size, so that the internal manifold is a lopsided product of unequal sub-manifolds, because that would violate the equations of motion: Maxwell's equation makes flux lines repel, so $T_{\mu\nu}$ would to be uniform in the extra dimensions, but ipso facto $R_{\mu\nu}$ would not be uniform in a lopsided vacuum.  It is therefore natural to conjecture that the endpoint of this instability is for the collapsing cycle to shrink to zero size and pinch off, causing spacetime to vanish in a bubble of nothing \cite{Witten:1981gj, Brown:2011gt}.  This process is akin to closed string tachyon condensation, described in \cite{Adams:2005rb}.


This `cycle-collapse instability' can  be understood in terms of the effective potential.  For example, if the internal manifold is $S_2\times S_2$, and you wrap $n$ units of a $4$-form uniformly around the entire internal manifold, and you fix the shape of both spheres, then the effective potential for the radii $R_1$ and $R_2$ is, schematically,
\begin{gather}
V_{\text{eff, }4\text{-form}}\sim\frac{1}{\left(R_1^{\;2}R_2^{\;2}\right)^{2/(p-2)}}\left( \frac{n^2}{\left(R_1^{\;2}R_2^{\;2}\right)^2} - \frac{1}{R_1^{\;2}}- \frac{1}{R_2^{\;2}}+\Lambda\right).
\end{gather}
The first term comes from the flux lines, which repel and push the radii out to larger values; the next two terms are from the curvatures of the spheres, which want the spheres to shrink to zero size.  The fact that the highest-form flux is insensitive to sub-curvatures is reflected in the fact that the flux term depends only on the total-volume combination $\left(R_1^{\;2}R_2^{\;2}\right)^2$.  The solution with $R_1=R_2$ sits at a saddle point of this effective potential; there is an instability for one sphere to expand while the other collapses and $V_\text{eff}\rightarrow-\infty$.

Because the `cycle-collapse instability' is related to the use of a highest-form flux, this suggests a simple solution to stabilize product manifolds: don't use a highest-form flux!  Instead of wrapping a single flux around the entire internal product manifold, wrap a lower-form flux individually around each element of the product.  
We can understand how this resolves the cycle-collapse instability from the perspective of the effective potential.  If you take the setup from above, but instead wrap $n_1$ units of a $2$-form flux around the first sphere, and $n_2$ units around the second sphere, the effective potential is
\begin{gather}
V_{\text{eff, }2\text{-form}}\sim\frac1{\left(R_1^{\;2}R_2^{\;2}\right)^{2/(p-2)}}\left( \frac{n_1^{\;2}}{R_1^{\;4}}+\frac{n_2^{\;2}}{R_2^{\;4}} - \frac{1}{R_1^{\;2}}- \frac{1}{R_2^{\;2}}+\Lambda\right).
\end{gather}
The flux term now depends on $R_1$ and $R_2$ individually, and is therefore sensitive to perturbations and able to restore the manifold to the vacuum.  In trying to stabilize the `cycle-collapse instability', we are led to consider Freund-Rubin compactifications with $N>1$.

\subsection{Results for General $N$}

No one likes having a movie spoiled,\footnote{It was his sled.} but the same is likely not true of technical physics papers: for $N>1$, we find that the `cycle-collapse instability' is essentially cured, that the other two classes of instability persist, and that no new classes arise.  

We find a `total-volume instability'  when $\Lambda>0$ and the flux density wrapping any sub-manifold gets too small  (as in the $N=1$ case).  

We find a `lumpiness instability' can exist when the right conditions are met.  For $N>1$:
\begin{itemize}
\item When $q=2$, all higher-mode fluctuations have positive mass.
\item When $q=3$, all higher-mode fluctuations have a stable mass squared.
\item When $q\ge4$, `lumpiness instabilities' can appear for any $\Lambda$.  This is unlike the $N=1$ case, where all compactifications with $\Lambda\le0$ were stable.  
\end{itemize}
These details are expanded upon and contrasted with the $N=1$ case in Figs.~2 and 3.
\newpage

 \begin{center}
 $\bm{N=1}$\\ \vspace{.12 in}
\begin{tabular}{|c||c|c|c|} \hline
  & $ \Lambda \leq 0$ &  $\Lambda > 0$  & $\Lambda>0$ \\  
   &  \hspace{5mm} {AdS}$_p$ minima \hspace{5mm} & \hspace{5mm}   {AdS}$_p$ minima \hspace{5mm} & \hspace{5mm}  {dS}$_p$ minima  \hspace{5mm} \ \\
\hline \hline
 &  &  & \\
  $q= 2$ & stable &  stable & stable \\
 &(always positive)  & (always positive)   &   \\
 \hline
 &  &  & \\
 $q=3$  &  stable &  stable & stable \\
  &  &  & \\
\hline
 &  &  & \\
 $q=4$  &  stable &  mostly unstable & mostly unstable \\
  &  &  (deep AdS$_p$ stable) & (high dS$_p$ stable) \\
\hline
 &  &  & \\
 $q=5, 7, 9 \ldots $  &  stable & unstable & unstable \\
  &  &  & \\
\hline
 &  &  & \\
 $q=6, 8, 10  \ldots $  &  stable &  mostly unstable & unstable \\
  &  &  (deep AdS$_p$ stable)  & \\
  \hline
\end{tabular}
\end{center}\vspace{-.19in}
\makebox[5.35in][r]{}\vspace{.11in}
\vspace{-.25in} 
\begin{figure}[htbp] 
   \caption{The stability of the shape modes of a single ($N=1$) $q$-sphere wrapped by a $q$-form flux, for  different signs of the higher-dimensional cosmological constant $\Lambda$.  AdS$_p$ solutions are always stable against zero-mode fluctuations and can exist for any sign of $\Lambda$; dS solutions, however, can only exist when $\Lambda>0$ and only some of them are stable to zero-modes.  The right-most column lists properties of those dS$_p$'s that are stable to zero-mode fluctuations, meaning they are minima of the effective potential in Fig.~1.  When $\Lambda\leq0$, the solution is always AdS and always stable to all fluctuations; though negative mass squareds exist for $q\geq 3$, they are always above the BF stability bound.  When $\Lambda>0$, the solution switches from AdS to Minkowski to dS depending on the flux and unstable mass squared can exist amongst the higher-mode fluctuations.  Stability in this case depends on $q$.  These results were derived in \cite{DeWolfe:2001nz, Bousso:2002fi} and we discuss them in Sec.~\ref{coupleddiagonalscalarsN1}.}
%
 %
  \label{fig:resultschart1}
\end{figure}


While we have essentially cured the `cycle-collapse instability', we find a residual version in the somewhat degenerate case when $M_{q,i}$ is itself a product.  
(For instance, consider the case where the internal manifold is $S_2\times S_2\times S_2\times S_2$.  If you wrap an 8-form flux around the whole thing, you're in the $N=1$ case and we've seen that you have an instability.  If you instead split the manifold up into two pairs of $S_2$'s, and wrap a 4-form flux individually around both pairs, then you're in the $N=2$ case, and you still have an instability.  Only in the $N=4$ case, where each sphere is individually wrapped by a 2-form flux, is the compactification stable.)  

There are no other instabilities beside these three: all of the extra types of fluctuations that exist when $N>1$, such as the angles between the sub-manifolds and the off-diagonal form fluctuations, all have positive mass.  We find the standard story for higher-spin fluctuations: a massless vector for every Killing vector of the internal manifold, a massless graviton, massless higher-spin form fields associated with harmonic forms of the internal manifold, and Kaluza-Klein towers of massive partners stacked above each of these massless fields.

 \begin{center}
  $\bm{N\ge2}$\\ \vspace{.12 in}
\begin{tabular}{|c||c|c|c|} \hline 
   & $ \Lambda \leq 0$ &  $\Lambda > 0$  & $\Lambda>0$ \\ 
   &  \hspace{5mm} {AdS}$_p$ minima \hspace{5mm} & \hspace{5mm}   {AdS}$_p$ minima \hspace{5mm} & \hspace{5mm}  {dS}$_p$ minima  \hspace{5mm} \ \\
\hline \hline
 &  &  & \\
  $q= 2$ & stable &  stable & stable \\
 &(always positive)  & (always positive)   &  \\
 \hline
 &  &  & \\
 $q=3$  &  stable &  stable & stable \\
  &  &  & \\
\hline
 &  &  & \\
 $q=4$  &  mostly unstable &  mostly unstable &  unstable \\
  & (some stable)  &  (some stable) & \\
\hline
 &  &  & \\
 $q=5, 7, 9 \ldots $  &  mostly unstable &  mostly unstable &  unstable \\
  & (some stable)  & (some stable)  &  \\
\hline
 &  &  & \\
 $q=6, 8, 10  \ldots $  &  mostly unstable &  mostly unstable &  unstable \\
  & (some stable)  & (some stable)  &  \\
  \hline
\end{tabular}
\end{center}\vspace{-.19in}
\makebox[5.35in][r]{}\vspace{.11in}
\vspace{-.25in} 
\begin{figure}[htbp] 
   %
   \caption{The stability of the shape modes of the product of $N$ $q$-spheres, each individually-wrapped by a $q$-form flux, for  different signs of the higher-dimensional cosmological constant $\Lambda$.  As in Fig.~2, the right-most column lists properties of those dS$_p$'s that are stable to zero-mode fluctuations, meaning they are minima of the effective potential in Fig.~1. 
   As in the $N=1$ case, when $\Lambda\leq0$, the solution is always AdS; and when $\Lambda>0$, the solution switches from AdS to Minkowski to dS depending on the flux.  Also as in the $N=1$ case, when the spheres each have $q=2$ or $q=3$ dimensions, the shape modes are always stable (for $q=3$  the mass squareds may be negative, but they are always greater than the BF stability bound). However, for higher $q$, there are differences from the $N=1$ case; most markedly, even for $\Lambda \leq 0$, the shape modes may be unstable. These results are derived in detail in Sec.~\ref{TGNCDSS}.}
%
   %
   %
   %
   %
  \label{fig:resultschart2}
\end{figure}


The results we find here refute 
claims in the literature.  In \cite{Kim:1985yq, Myung:1987ec, Myung:1986da}, Minkowksi compactifications of the form $M_4\times S_2\times S_2$ and $M_4\times S_2\times S_2\times S_2$ were argued to be unstable to $\ell=1$ perturbations; below we show how correct handling of residual gauge invariance proves these modes, and indeed all modes of these compactifications, are completely stable.  
%
%
%
%
%
%

Our results encompass and generalize previous work on the $N=1$ case in \cite{DeWolfe:2001nz,Bousso:2002fi}.

\subsection{Notation}
We will be investigating product manifolds of the form $M_p\times M_{q,1} \times\cdots\times M_{q,N}$.   We use coordinate $x$ and indices $\mu$, $\nu$, $\dots$ for the $M_p$ and coordinates $y_i$ and indices $\alpha_i$, $\beta_i$, $\dots$ for $M_{q,i}$.  Capital Roman indices $M$, $N$, $\dots$ run over the whole manifold. We define the exterior derivative as $(dw)_{\alpha_1\cdots\alpha_k}=k\nabla_{[\alpha_1}w_{\alpha_2\cdots\alpha_k]}$ and the exterior co-derivatives as $(d^{\dagger}w)_{\alpha_3\cdots\alpha_k}=\nabla^{\alpha_2}w_{\alpha_2\cdots\alpha_k}$.  If $k$ indices are enclosed in square brackets $[\dots]$, they are antisymmetrized over and a combinatoric factor $1/k!$ is included.  If $k$ indices are enclosed in parentheses $(\dots)$, they are symmetrized over, a combinatoric factor $1/k!$ is included, \emph{and the trace is removed}.  This final part not being the norm, we have added periodic reminders throughout the text.  The Riemann tensor is defined so that $2\nabla_{[A}\nabla_{B]}V_C=-R^D_{\;\;CAB}V_D$ and $R_{AB}=R^C_{\;\;ACB}$. We use a mostly plus metric signature $(-,+,\dots,+)$, which means that the scalar Laplacian $\Box$ and the Hodge Laplacian $\triangle=(d+d^\dagger)^2$ are both negative semi-definite.

\section{Background Solution}

We will investigate compactifications of $D=p+N q$ dimensions down to $p$ dimensions, where the internal manifold is the product $M_{q,1} \times\cdots\times M_{q,N}$, and a $q$-form flux $F_q$ wraps each $M_{q,i}$ individually.  These compactifications will be solutions to the equations of motion that follow from the action
\begin{equation}
S=\int d^pxd^{q}y_1\cdots d^{q}y_N \sqrt{-g}\left(R-\frac1{2q!}F_q^{\;2}-2\Lambda\right),
\end{equation}
where $\Lambda$ is a higher-dimensional cosmological constant.

Einstein's equation is
\begin{equation}
\label{0Einstein}
R_{MN}=\bar{T}_{MN}\equiv\frac12\frac1{(q-1)!}F_{MP_2\cdots P_q}F_N^{\;\;\;P_2\cdots P_q}-\frac12\frac{q-1}{D-2}\frac1{q!}F_q^{\;2}g_{MN}+\frac2{D-2}\Lambda g_{MN},
\end{equation}
where $\bar{T}_{MN}$ is the trace-subtracted energy momentum tensor $\bar{T}_{MN}\equiv T_{MN}-\frac{1}{D-2} T_P^{\;\;P} g_{MN}$.  Maxwell's equation is 
\begin{equation}
\label{0Maxwell}
d^{\dagger} F_q\equiv\nabla^MF_{M P_2...P_q}=0.
\end{equation}
The $q$-form flux $F_q$ is the exterior derivative of a flux potential $A_{q-1}$, so $F_q=dA_{q-1}$.

We look for solutions where the $q$-form flux wraps each $M_q$ separately 
\begin{equation}
F_q=\sum_{i=1}^N c_i \; \text{vol}_{M_{q,i}},
\end{equation}
where $c_i$ is the flux density  and $\text{vol}_{M_{q,i}}$ is the volume form for the $i$th  $M_q$; $\text{vol}_{M_{q,i}}$ is only non-zero when all $q$ indices are from $M_{q,i}$, in which case it is equal to the $q$-dimensional Levi-Civita tensor density.
This ansatz automatically solves Maxwell's equation Eq.~\eqref{0Maxwell}; for compactifications without warping, Maxwell's equation demands that the $q$-form flux is uniform in the $q$-cycle it wraps.

That the flux is uniform forces the $M_{q,i}$ to be Einstein, which in turn guarantees that the extended dimensions $M_p$ are maximally symmetric.  We define radii of curvature $L$ and $R_i$ for $M_p$ and $M_{q,i}$, respectively, so that
\begin{gather}
R_{\mu\nu}=\frac{p-1}{L^2}g_{\mu\nu}, \hspace{.4 in} R_{\alpha_i\beta_j}=\frac{q-1}{R_i^{\;2}}g_{\alpha_i\beta_j} \delta_{ij}, \hspace{.4 in} R_{\mu\alpha_i}=0.
\end{gather}
All the off-diagonal terms are 0.  The $R_i$ are all positive, but we allow for analytic continuation to imaginary $L$.  When $L^{-2}>0$, $M_p$ is a de Sitter space; when $L^{-2}<0$, $M_p$ is an Anti-de Sitter space; and when $L^{-2}=0$, $M_p$ is a Minkowski space.

Einstein's equation Eq.~\eqref{0Einstein} relates the distance scales to the flux densities
\begin{eqnarray}
\label{0sol1}
\frac{p-1}{L^2}&=&-\frac12\frac{q-1}{D-2}\sum_{i=1}^N c_i^{\;2} + \frac2{D-2}\Lambda  \\
\frac{q-1}{R_i^{\;2}}&=&\frac12c_i^{\;2}+\frac{p-1}{L^2}.
\label{0sol2}
\end{eqnarray}
The solution is Minkowski ($L^{-2}=0$) when the $c_i$ satisfy
\begin{gather}
\label{MinkowskiCondition}
\sum_{i=1}^N c_i^{\;2}=\frac{4}{q-1}\Lambda.
\end{gather}
The solution with no flux at all ($c_i=0$) is sometimes called the Nariai solution; the solution where one of the $c_i\rightarrow\infty$, sending $R_i\rightarrow0$ and $L^{-2}\rightarrow-\infty$, we argued in \cite{Brown:2011gt}, should be thought of as the `nothing state'.

Equations \eqref{0sol1} and \eqref{0sol2} provide a solution to the equations of motion, but we do not yet know if this solution is stable or unstable.  To determine its stability, we need to find the mass spectrum of small fluctuations around this background.  For de Sitter or Minkowski compactifications, stability means that all of these masses are positive; for AdS compactifications, the mass squareds may be negative---stability means that all of the mass squareds are no more negative than the BF bound \cite{Breitenlohner:1982bm}.

To first order, the equations of motion for small fluctuations around the background solution are coupled partial differential equations.  Finding the spectrum means solving these equations, and that will be the project of the bulk of this paper.  We will solve them in two steps.  First, we perform two simultaneous decompositions on the fluctuations: a Hodge decomposition into transverse and longitudinal parts, and a decomposition into eigenvectors of the Lichnerowicz Laplacian.  These two decompositions break the coupled partial differential equations apart into coupled ordinary differential equations.  Second, by choosing the right combinations of fields, we can diagonalize the equations, and from the resulting decoupled ordinary differential equations, we can directly read off the spectrum.  
In Sec.~\ref{sec:EOMs}, we will derive the first-order equations of motion.  Step one of the solution happens in Sec.~\ref{sec:decomposingflesh} and step two happens in Sec.~\ref{sec:diagonalizing}.

\section{First-Order Equations of Motion}
\label{sec:EOMs}

Consider small perturbations to the background fields $g_{MN}$ and $A_{P_2\cdots P_q}$:
\begin{equation}
g_{MN}\rightarrow g_{MN}+h_{MN}, \hspace{.3 in} \text{and}\hspace{.3 in} A_{P_2\cdots P_q}\rightarrow A_{P_2\cdots P_q}+B_{P_2\cdots P_q}.
\end{equation}
We will define $f_q=dB_{q-1}$ so that 
\begin{equation}
F_q=dA_q\rightarrow F_q+f_q.
\end{equation}

To first order in the fluctuations $h_{MN}$ and $B_{P_2\cdots P_q}$, Einstein's equation becomes
\begin{equation}
\label{Einstein1}
R_{MN}^{\;(1)}=\bar{T}_{MN}^{\;(1)},
\end{equation}
where
\begin{gather}
R_{MN}^{\;(1)}=-\frac12\Big[\Box h_{MN}+\nabla_M\nabla_Nh^P_{\;\;P} -\nabla_M\nabla^Ph_{PN}-\nabla_N\nabla^Ph_{PM} \nonumber \\
\label{R1}
-2R_{M\;\;\;\;N}^{\;\;\;PQ}h_{PQ} -R_M^{\;\;\;P}h_{PN}-R_N^{\;\;\;P}h_{PM}\Big],
\end{gather}
and
\begin{gather}
\label{T1}
\bar{T}_{MN}^{\;(1)}=-\frac12\frac1{(q-2)!}F_{M\;\;P_3\cdots P_q}^{\;\;\;Q}F_{N}^{\;\;RP_3\cdots P_q}h_{QR}+\frac12\frac{q-1}{D-2}\frac1{(q-1)!}\left(F_{\;\;P_2\cdots P_q}^{Q}F^{RP_2\cdots P_q} h_{QR}\right)g_{MN} \nonumber\\
+\frac{p-1}{L^2}h_{MN}+\frac12\frac1{(q-1)!}\left(f_{MP_2\cdots P_q}F_N^{\;\;\;P_2\cdots P_q}+f_{NP_2\cdots P_q}F_M^{\;\;\;P_2\cdots P_q}\right)\nonumber \\
-\frac{q-1}{D-2}\frac1{q!}(f_{P_1\cdots P_q} F^{P_1\cdots P_q}) g_{MN}.
\end{gather}

To first order in the fluctuations $h_{MN}$ and $B_{P_2\cdots P_q}$, Maxwell's equation becomes
\begin{equation}
\label{Maxwell1}
\nabla^Mf_{MP_2\cdots P_q}-g^{MN}\Gamma_{MN}^{Q\;(1)}F_{QP_2\cdots P_q}-\sum_{k=1}^q\Gamma_{MP_k}^{Q\;(1)}F^{M}_{\;\;\;P_2\cdots P_{k-1}QP_{k+1}\cdots P_q}=0,
\end{equation}
where 
\begin{equation}
\Gamma_{MN}^{P\;(1)}=\frac12\left(\nabla_M h_N^{\;\;\;P}+\nabla_N h_M^{\;\;\;P}-\nabla^P h_{MN}\right)
\end{equation}
is the Christoffel symbol to first order in $h_{MN}$.  Notice that if at least two of the indices $P_2$, \dots, $P_q$ come from different sub-manifolds then only the first term in Eq.~\eqref{Maxwell1} is non-zero; in other words, more than singly off-diagonal terms in Maxwell's equation decouple from gravity.
In order to determine the stability and spectrum of these compactifications, we need to solve Eqs.~\eqref{Einstein1} and \eqref{Maxwell1}, which are coupled partial differential equations.  That project begins in Sec.~\ref{sec:decomposingflesh}.

\section{Decomposing the Fluctuations}
\label{sec:decomposingflesh} 
The Lichnerowicz operator $\triangle_L$ is a generalization of the Laplacian to tensors; it is given by
\begin{equation}
\triangle_L T_{a_1...a_m}\equiv\Box T_{a_1...a_m}-\sum_{i=1}^m R^c_{\;a_i}T_{a_1\cdots a_{i-1}ca_{i+1}\cdots a_m}+\sum_{i,j=1, i\ne j}^{i,j=m} R^{c\;\;d}_{\;a_i\;a_j}T_{a_1\cdots a_{i-1}ca_{i+1}\cdots a_{j-1}da_{j+1}\cdots a_m}.
\end{equation}
Acting on a scalar, the Lichnerowicz operator is the Laplacian $\triangle_L T = \Box T$.  Acting on a vector, it is the Laplacian shifted by a term proportional to the Ricci tensor  $\triangle_L T_a = \Box T_a-R^{b}_{\;a}T_b$.  Acting on a symmetric 2-tensor, it is the Laplacian shifted by terms proportional to the Ricci and Riemann tensors $\triangle_L T_{ab} = \Box T_{ab}-R^{c}_{\;a}T_{cb}-R^{c}_{\;b}T_{ca}+2R^{c\;d}_{\;a\;\;b}T_{cd}$.  Finally, acting on a differential $k$-form, the Lichnerowicz operator is equal to the Hodge Laplacian $\triangle_L F_k =(d+d^\dagger)^2F_k$.  On an Einstein manifold, $\triangle_L$ commutes with traces, gradients, and symmetrized derivatives.

The reason this is a helpful definition is clear from Eq.~\eqref{R1}: all the curvature terms can be collected into a single Lichnerowicz  Laplacian. Eigenvectors of $\triangle_L$, therefore, aren't coupled by $R_{MN}$; if they are coupled it is only by $T_{MN}$.  The equations of motion also decouple under the Hodge decomposition into longitudinal and transverse.

Because our internal manifold is a product, we have a choice: we can either decompose on the whole manifold at once, or we can decompose separately on each element of the product.  The equations of motion make the choice for us---only in the second case does $T_{MN}$ decouple.  This choice, however, introduces some new complications for the Hodge decomposition that we would like to be conducting simultaneously.  In Sec.~\ref{Lichnerowitcz}, we discuss the simultaneous Hodge and Lichnerowicz decomposition and address this complication; this section also serves to define notation used in the rest of the paper.  In Sec.~\ref{decomp}, we give the decomposition explicitly for our fluctuation fields, and in Sec.~\ref{eom1} we plug in to the equations of motion.

\subsection{The Lichnerowicz/Hodge Decomposion}
\label{Lichnerowitcz}

It will be helpful to discuss a few simple examples explicitly first, before we return to our fluctuation fields. In this subsection, let's forget about extended dimensions and consider decomposing on a compact Einstein manifold $M_{Nq}$.  Below,  we will discuss the Hodge/Lichnerowicz decomposition of scalar fields, vector fields, symmetric two-tensors, and differential $k$-forms.  In each case, we first discuss the decomposition generally  on a compact Einstein manifold $M_{q}$, and then we give special attention to the case of a product manifold $M_{Nq}=M_{q,1}\times\cdots\times M_{q,N}$.

\textbf{Scalars:}  We will denote scalar eigenvectors by $Y^I$ and their eigenvalues by $\lambda^I$:
\begin{equation}
\triangle_L Y^{I}\equiv\Box Y^{I}=\lambda^{I} Y^{I}.
\end{equation}
When the internal manifold is a $q$-sphere, $\lambda^I=-\ell(\ell+q-1)/R^2$, where $\ell = 0, 1, 2, \dots$ is the angular momentum and $R$ is the radius of the sphere.
Eigenvectors of the Laplacian on a compact manifold give a complete basis, so a general scalar field $\phi(y)$ can be decomposed
\begin{equation}
\phi(y)=\sum_I \phi^I Y^I(y),
\end{equation}
where $\phi^I$ denotes the component of $\phi(y)$ that lies along the eigenvector $Y^I$.
The scalar Laplacian on a compact manifold is negative semi-definite: all the $\lambda^I\le0$.  In fact, Lichnerowicz \cite{Lichnerowicz} proved that there is only a single zero eigenvalue $\lambda^{I=0}=0$, and that the next least negative eigenvalue is bounded from above: $\lambda^{I\ne0}\le-q/R^2$, where $R$ is the radius of curvature.  The bound is saturated when $Y=Y^{I=C}$ is a conformal scalar, which satisfies
\begin{gather}
\label{ConformalScalar}
\nabla_{(\alpha}\nabla_{\beta)}Y^{I=C}\equiv\nabla_\alpha\nabla_\beta Y^{I=C}-\frac1q g_{\alpha\beta}\Box Y^{I=C}=0.
\end{gather} 
Small diffeomorphisms along the direction $\nabla_\alpha Y^{I=C}$ only affect the conformal factor of the metric.  When the internal manifold is a $q$-sphere, the $\ell=1$ mode is exactly such a conformal scalar;  indeed conformal scalars \emph{only exist} when the internal manifold is a sphere \cite{Obata}.  Taking divergence of Eq.~\eqref{ConformalScalar} proves that $\lambda^{I=C}$ indeed saturates the Lichnerowicz bound.

When the internal manifold is a product, we can be more specific.  The Laplacian breaks up into pieces on each sub-manifold $\Box_y=\Box_{y_1}+\cdots+\Box_{y_N}$, so we can write the eigenvectors and eigenvalues for the full manifold explicitly  in terms of the eigenvectors and eigenvalues on each sub-manifold:
\begin{gather}
Y(y)^I=Y_1^{I_1}(y_1)\cdots Y_N^{I_N}(y_N),\hspace{.3in}\text{and}\hspace{.3in}\lambda^I=\sum_{k=1}^N\lambda_k^{I_k}.
\end{gather}
We will use $Y^I$ as a shorthand for this product.  When the internal manifold is a product of $N$ $q$-spheres, each eigenvector is identified by a list of $N$ spherical harmonic $\ell$'s.

\textbf{Vectors:}
The Hodge decomposition theorem states that vector fields can be uniquely decomposed into a transverse and a longitudinal part, $V_\alpha(y)=V^T_\alpha(y)+ V^L_\alpha(y)$, with $\nabla^\alpha V^T_\alpha=0$ and $V^L_\alpha(y)$ expressible as the divergence of a scalar $V^L_\alpha(y)=\nabla_\alpha\phi(y)$.  We will denote transverse vector eigenvectors by $Y_\alpha^{I}$ and their eigenvalues by $\kappa^I$:
\begin{equation}
\triangle_L Y_\alpha^I\equiv\Box Y_\alpha^I-R_\alpha^{\;\;\beta}Y_\beta^I=\kappa^I Y_\alpha^I,
\end{equation}
where $\nabla^{\alpha}Y_\alpha^I=0$.  When the internal manifold is a $q$-sphere $\kappa^I=-(\ell+1)(\ell+q-2)/R^2$, where $\ell=$ 1, 2, $\dots$ is the angular momentum.  The $Y_\alpha^I$ form a complete basis for transverse vectors.
Gradients of the scalar eigenvectors satisfy
\begin{gather}
\triangle_L \nabla_\alpha Y^I = \lambda^I  \nabla_\alpha Y^I,
\end{gather}
and form a basis for longitudinal vectors.  A general vector field $V_\alpha(y)$ can therefore uniquely be decomposed as
\begin{gather}
V_\alpha(y)=\sum_I V^{T,I} Y_\alpha^I+V^{L,I} \nabla_\alpha Y^I.
\end{gather}
The divergence of $V_\alpha$ is
\begin{gather}
\nabla^\alpha V_\alpha(y)=\sum_I V^{L,I} \lambda^I Y^I;
\end{gather}
because the $Y^I$ form an orthonormal basis (once we've defined a reasonable dot product), $\nabla^\alpha V_\alpha(y)=0$ if and only if $V^{L,I}=0$ for all $I$.

Similarly to the scalar case, $\triangle_L$ is negative definite.  In this case, there is no zero mode and all the eigenvalues are bounded from above by a Lichnerowicz bound $\kappa^I\le-2(q-1)/R^2$.  The inequality is saturated when $Y_\alpha=Y_\alpha^{I=K}$ is a Killing vector, which satisfies
\begin{gather}
\nabla^{}_\alpha Y_\beta^{I=K}+\nabla^{}_\beta Y^{I=K}_\alpha=0.
\label{KillingVec}
\end{gather}
Small diffeomorphisms along the direction $Y_\alpha^{I=K}$ leave the metric invariant. On a sphere, the $\ell = 1$ vector spherical harmonic is a Killing vector.  Taking the trace and divergence of Eq.~\eqref{KillingVec} proves that $Y^{I=K}$ is transverse and that $\lambda^{I=K}$ indeed saturates the Lichnerowicz bound.

In the case that the compact manifold is a product manifold, the eigenvectors on the full manifold can be written as a product of the eigenvectors on each sub-manifold.  If the vector's single index comes from $M_{q,i}$, then all the other $M_{q,j}$'s contribute a scalar eigenvector to the product and $M_{q,i}$ contributes a vector eigenvector, either transverse or longitudinal.  If it's transverse, we define the shorthand
\begin{gather}
Y_{\alpha_i}^I(y)\equiv Y^{I_1}(y_1)\cdots Y^{I_{i-1}}(y_{i-1})Y^{I_i}_{\alpha_i}(y_i)Y^{I_{i+1}}(y_{i+1})\cdots Y^{I_N}(y_N),
\end{gather}
and if it's longitudinal, we can write it as the product as
\begin{gather}
\nabla_{\alpha_i}^I Y(y)= Y^{I_1}(y_1)\cdots Y^{I_{i-1}}(y_{i-1})\nabla_{\alpha_i}Y^{I_i}(y_i)Y^{I_{i+1}}(y_{i+1})\cdots Y^{I_N}(y_N).
\end{gather}
These two together form a complete basis for vectors fields, so a general vector field can be decomposed as 
\begin{gather}
V_{\alpha_i}(y)=\sum_I V_i^{T,I} Y^I_{\alpha_i}(y) + V_i^{L,I}\nabla_{\alpha_i} Y^I(y).
\end{gather}
The associated eigenvalues are
\begin{gather}
\triangle_L Y_{\alpha_i}^I=\left(\sum_{k\ne i}\lambda_k^{I_k}+\kappa_i^{I_i}\right) Y_{\alpha_i},\hspace{.3in} \text{   and   } \hspace{.3in}\triangle_L\nabla_{\alpha_i}Y=\left(\sum_{k}\lambda_k^{I_k}\right) \nabla_{\alpha_i}Y.
\end{gather}

We have broken the vector into $N$ transverse parts $V_k^{T,I}$, and $N$ longitudinal parts $V_k^{L,I}$.  Crucially, despite the fact that the $V_k^{L,I}$ refer to components that are longitudinal on a $M_{q,k}$, we will be able to form combinations that are transverse on the whole manifold.  To see this, try setting the divergence of $V_\alpha$ to zero: 
\begin{gather}
\sum_{k=1}^N\nabla^{\alpha_k}V_{\alpha_k}=\sum_I \sum_{k=1}^N \lambda_k^I V_k^{L,I}Y^I(y)=0.
\end{gather}
Because the $Y^I(y)$ are orthonormal, we see that enforcing transversality of $V_\alpha$ enforces one condition on the $V_k^{K,I}$ for each $I$:
\begin{gather}
\sum_{k=1}^N\lambda_k^{I} V_k^{L,I}=0.
\end{gather}
Of the $N$ independent modes $V_k^{L,I}$, therefore, we can construct $N-1$ linearly independent combinations that are transverse on the whole manifold; they are not transverse on a given sub-manifold, but their sub-divergences cancel for the whole manifold.  In total, therefore, we have broken the vector up into $2N$ components, of which $2N-1$ are transverse and $1$ is longitudinal.

\textbf{Symmetric 2-tensors:} The Hodge decomposition theorem states that symmetric 2-tensors can be uniquely decomposed into a trace, a transverse traceless tensor, and a component proportional to the divergence of a vector; that vector can then further be decomposed into a transverse and longitudinal part:  $T_{\alpha\beta}(y)=T_\gamma^{\;\gamma}(y)g_{\alpha\beta} + T_{(\alpha\beta)}^{TT}(y)+\nabla^{\;}_{(\alpha}V^T_{\beta)}(y) + \nabla_{(\alpha}\nabla_{\beta)}\phi(y)$.\footnote{As a quick reminder, subscripts in parentheses mean both symmetrized and trace-free.  In particular, because $V^T_\beta$ is transverse, $\nabla^{}_{(\alpha}V^T_{\beta)}=\frac12\left(\nabla^{}_{\alpha}V^T_{\beta}+\nabla^{}_{\beta}V^T_{\alpha}\right)$ and $\nabla_{(\alpha}\nabla_{\beta)}\phi=\nabla_\alpha\nabla_\beta\phi-\frac1q\Box\phi\;g_{\alpha\beta}$.}  The superscript $T$'s indicate transversality, so $\nabla^\alpha T^{TT}_{(\alpha\beta)}=\nabla^\beta T^{TT}_{(\alpha\beta)}=0$ and $\nabla^\beta V_\beta^T=0$.

We will denote transverse traceless tensor eigenvectors of the Lichnerowicz Laplacian by $Y_{(\alpha\beta)}^{I}$ and their eigenvalues by $\tau^I$:
\begin{equation}
\triangle_L Y_{(\alpha\beta)}^{I}=\tau^I Y_{(\alpha\beta)},
\end{equation}
and these form a complete basis for transverse traceless symmetric 2-tensors. If the internal manifold is a $q$-sphere, then $\tau^I=-[\ell(\ell+q-1)+2(q-1)]/R^2$ where $\ell=2, 3, \dots$ is the angular momentum.  Unlike scalar and vector eigenvalues, the $\tau^I$ do not necessarily satisfy a Lichnerowicz bound.  If the internal manifold is topologically a sphere, then $\tau^I\le-4q/R^2$, which is saturated by the $\ell=2$ mode of the sphere.  However, if the internal manifold is a product space, $\tau^I$ can violate this bound.  For instance, the mode in which one sub-manifold swells while the rest shrinks in a volume-preserving way has eigenvalue $\tau^I=0$; the absence of a Lichnerowicz bound is connected to the `cycle-collapse instability' discussed in Sec.~\ref{N1}.

The trace is decomposed as a scalar, so let's focus on the traceless part.  A symmetric traceless tensor can be uniquely decomposed as
\begin{gather}
\label{Tdecom}
T_{(\alpha\beta)}(y)=\sum_I T^{TT,I} Y_{(\alpha\beta)}^I(y) + T^{LT,I} \nabla_{(\alpha}Y_{\beta)}^I(y) + T^{LL,I} \nabla_{(\alpha}\nabla_{\beta)} Y^I(y),
\end{gather}
where
\begin{gather}
\triangle_L\nabla^{}_{(\alpha} Y_{\beta)}^I=\kappa^I \nabla^{}_{(\alpha} Y_{\beta)}^I,\hspace{.3in}\text{and}\hspace{.3in}\triangle_L\nabla_{(\alpha} \nabla_{\beta)} Y^I = \lambda^I\nabla_{(\alpha} \nabla_{\beta)}Y^I. 
\end{gather}
The divergence of $T_{(\alpha\beta)}$ is
\begin{gather}
\nabla^\alpha T_{(\alpha\beta)}=\sum_I T^{LT,I}\left(\frac12\kappa^I+\frac{q-1}{R^{\;2}}\right)Y_\beta^I + T^{LL,I} \left(\frac{q-1}q\lambda^I+\frac{q-1}{R^{\;2}}\right) \nabla_\beta Y^I,
\end{gather}
where we used the fact that the internal manifold is Einstein, $R_{\alpha\beta}=(q-1)/R^2 \; g_{\alpha\beta}$.
Because the $Y_\beta^{I}$ and $\nabla_\beta Y^I$ form an orthonormal basis, setting the divergence to zero requires setting each term in the sum individually to zero; for each $I$, either the $T$'s must be zero or the terms in parentheses must be zero.  But the terms in parentheses are only zero if a Lichnerowicz bound is saturated, and Killing vectors and conformal scalars don't contribute to the sums in Eq.~\eqref{Tdecom}.  Therefore, $\nabla^\alpha T_{(\alpha\beta)}=0$ if and only if $T^{LT,I}=T^{LL,I}=0$ for all $I$.  

As before, if the internal manifold is a product of Einstein spaces, we can write the eigenvectors on the full manifold as products of eigenvectors on each sub-manifold.  We will need to distinguish between diagonal blocks, where both indices come from the same sub-manifold, and off-diagonal blocks, where the indices come from different sub-manifolds.  For diagonal blocks, we define the shorthand
\begin{gather}
Y_{(\alpha_i\beta_i)}^I(y)\equiv Y^{I_1}(y_1)\cdots Y^{I_{i-1}}(y_{i-1})Y^{I_i}_{(\alpha_i\beta_i)}(y_i)Y^{I_{i+1}}(y_{i+1})\cdots Y^{I_N}(y_N),
\end{gather}
and for the off-diagonal blocks ($i\ne j$), we define the shorthand
\begin{gather}
\nonumber
Y_{\alpha_i\beta_j}^I(y)\equiv Y^{I_1}(y_1)\cdots Y^{I_{i-1}}(y_{i-1})Y^{I_{i}}_{\alpha_i}(y_i)Y^{I_{i+1}}(y_{i+1})\cdots Y^{I_{j-1}}(y_{j-1})Y^{I_j}_{\beta_j}(y_j)Y^{I_{j+1}}(y_{j+1})\cdots Y^{I_N}(y_N).
\end{gather}

A symmetric tensor field on a product manifold can therefore be uniquely decomposed as
\begin{eqnarray}
T_{\alpha_i\beta_i}&=&\sum_I T_i^I g_{\alpha\beta} Y^I + T^{TT,I}_i Y_{(\alpha_i\beta_i)}^I + T^{LT,I}_i \nabla_{(\alpha_i}Y_{\beta_i)}^I + T^{LL,I}_i \nabla_{(\alpha_i}\nabla_{\beta_i)} Y^I \nonumber\\
T_{\alpha_i\beta_j}&=&\sum_I T^{TT,I}_{ij} Y_{\alpha_i\beta_j}^I + T^{LT,I}_{ij} \nabla^{}_{\alpha_i}Y_{\beta_j}^I + T^{TL,I}_{ij} \nabla^{}_{\beta_j}Y_{\alpha_i}^I+T^{LL,I}_{ij} \nabla_{(\alpha_i}\nabla_{\beta_j)} Y^I, 
\end{eqnarray}
where the first line is for the diagonal blocks and the second line is for off-diagonal blocks ($i\ne j$).  Decomposing the off-diagonal blocks is akin to squaring the decomposition of a vector---each index contributes a vector harmonic, either $T$ or $L$.  Symmetry of $T_{\alpha\beta}$ imposes the constraints
\begin{equation}
T_{ij}^{TT,I}=T_{ji}^{TT,I}, \hspace{1.3cm}  T_{ij}^{LT,I}=T_{ji}^{TL,I}, \hspace{1.3cm} T_{ij}^{LL,I}=T_{ji}^{LL,I}.
\end{equation}
The associated eigenvalues are:
\begin{gather}
\triangle_L Y_{(\alpha_i\beta_i)}^I=\left(\sum_{k\ne i}\lambda_k^{I_k}+\tau_i^{I_i}\right) Y_{(\alpha_i\beta_i)}^I,\hspace{.3in}  \triangle_L\nabla_{(\alpha_i}Y_{\beta_i)}^I=\left(\sum_{k\ne i}\lambda_k^{I_k} + \kappa_i^{I_i}\right) \nabla_{(\alpha_i}Y_{\beta_i)}^I, \nonumber\\
\triangle_L \nabla_{(\alpha_i}\nabla_{\beta_i)} Y^I = \left(\sum_{k}\lambda_k^{I_k}\right)\nabla_{(\alpha_i}\nabla_{\beta_i)} Y^I, \nonumber\\\
\triangle_L Y_{\alpha_i\beta_j}^I=\left(\sum_{k\ne i,j}\lambda_k^{I_k}+\kappa_i^{I_i}+\kappa_j^{I_j}\right) Y_{\alpha_i\beta_j}^I, \hspace{.3in} 
\triangle_L\nabla_{\alpha_i}Y_{\beta_j}^I=\left(\sum_{k\ne j}\lambda_k^{I_k} + \kappa_j^{I_j}\right) \nabla_{\alpha_i}Y_{\beta_j}^I, \nonumber\\
\text{and}\hspace{.3in} \triangle_L \nabla_{\alpha_i}\nabla_{\beta_j} Y^I = \left(\sum_{k}\lambda_k^{I_k}\right)\nabla_{\alpha_i}\nabla_{\beta_j} Y^I.
\end{gather}

In breaking up the tensor field $T_{\alpha\beta}$ like this, we have identified $N$ sub-traces $T_i$, one from each diagonal block.  Because the background solution is not uniform over the whole internal manifold, just over the sub-manifolds individually, it will be helpful to think of `tracelessness' as subtracting off all $N$ sub-traces, rather than just subtracting off the total trace.  

As before, we can find combinations of longitudinal sub-parts that are transverse on the whole internal manifold.  To see this, take the divergence of the part that remains after removing the $N$ sub-traces:
\begin{gather}
\sum_{k\ne i}\nabla^{\alpha_k}T_{\alpha_k\beta_i}+\nabla^{\alpha_i}T_{(\alpha_i\beta_i)}=\sum_I \left[\sum_{k\ne i} \lambda_i T_{ki}^{LT,I}+\left(\frac12\kappa_i+\frac{q-1}{R_i^{\;2}}\right) T_i^{TL,I}\right] Y_{\beta_i}^I \nonumber \\
+\left[\sum_{k\ne i} \lambda_i T_{ki}^{LL,I}+\left(\frac{q-1}q\lambda_i+\frac{q-1}{R_i^{\;2}}\right) T_i^{LL,I}\right]\nabla_{\beta_i}Y^I=0.
\end{gather}

\newpage

Because the eigenbasis is orthonormal, enforcing transversality of the traceless part of $T_{\alpha\beta}$ is equivalent to enforcing two conditions for each $I$:
\begin{gather}
\sum_{k\ne i} \lambda_k T_{ki}^{LT,I}=-\left(\frac12\kappa_i+\frac{q-1}{R_i^{\;2}}\right) T_i^{TL,I},\nonumber\\\text{and}\hspace{.3in}\sum_{k\ne i}^N \lambda_k T_{ki}^{LL,I}=-\left(\frac{q-1}q\lambda_i+\frac{q-1}{R_i^{\;2}}\right) T_i^{LL,I}.
\end{gather}

In total, a symmetric two-tensor is broken up into $N$ sub-traces, $3 \times N$ diagonal-block traceless tensors, and $4\times N(N-1)/2$ off-diagonal tensors; transversality enforces $2N$ conditions.

\textbf{Differential k-forms:} The Hodge decomposition theorem states that differential form fields can also be broken up into a transverse and longitudinal part, and that the transverse part can be further broken up into a co-exact and a harmonic part.  We will use superscripts $T$ for co-exact, $L$ for longitudinal, and $H$ for harmonic.
Define co-exact eigenvectors and eigenvalues of the Lichnerowicz Laplacian, which for forms equals the Hodge Laplacian, as
\begin{gather}
\left[(d+d^\dagger)^2Y_k\right]_{[\alpha_1\cdots\alpha_k]}=\triangle_L Y_{[\alpha_1\cdots\alpha_k]}=\tau^{(k)} Y_{[\alpha_1\cdots\alpha_k]}\newline
\end{gather}
where $d^\dagger Y_q=0$.  For $q$-spheres,  $\tau^{(k)}=-[(\ell+k)(\ell+q-1-k)-k+k^2]/R^2$.  Because they are co-exact, our $Y_k$ can be written as 
\begin{gather}
Y_{[\alpha_1\cdots\alpha_k]}=\epsilon^{\gamma_1\cdots\gamma_{q-k}}_{\;\;\;\;\;\;\;\;\;\;\;\;\;\;\;\alpha_1\cdots\alpha_k}\nabla_{\gamma_1} Y_{[\gamma_2\cdots\gamma_{q-k}]}.
\end{gather}
In other words, we use the fact that $d_q \star_q Y_k=0$ to write  $Y_k=\star_qd_q Y_{q-k-1}$, a useful  relation for forms with $k>q/2$.  Harmonic eigenvectors satisfy
\begin{gather}
d Y^H_k=d^{\dagger} Y^H_k=(d+d^\dagger)^2Y^H_k=0.
\end{gather}
Harmonic eigenvectors have zero eigenvalue; co-exact eigenvalues are all bounded from above.

A differential $k$-form field can be uniquely decomposed as
\begin{gather}
F_{[\alpha_1\cdots\alpha_k]}(y)=\sum_I F^{T,I} Y_{[\alpha_1\cdots\alpha_k]}^I(y) + F^{L,I} \nabla^{}_{[\alpha_1}Y_{\alpha_2\cdots\alpha_k]}^I(y) + F^{H,I} Y_{[\alpha_1\cdots\alpha_k]}^{H,I}(y).
\end{gather}

As before, if the manifold is a product, we can write the eigenvectors explicitly as a product of eigenvectors on the sub-manifolds.  The $k$ indices of a $k$-form are distributed amongst the $N$ sub-manifolds; each sub-manifold contributes an eigenvector with the appropriate number of indices to the product, either $T$ or $L$ or $H$.  We will write these products in our usual shorthand, so that, for instance, the component  of a 4-form that has 2 longitudinal indices on the $i$th manifold, 2 transverse indices on the $j$th, and none on the rest, will be written as:
\begin{gather}
\nabla^{}_{[\alpha_{1,i}}Y^I_{\alpha_{2,i}][\beta_{1,j}\beta_{2,j}]}\equiv \nabla^{\;_{}}_{[\alpha_{1,i}}Y^{I_i}_{\alpha_{2,i}]}(y_i)Y^{I_j}_{[\beta_{1,j}\beta_{2,j}]}(y_j) \prod_{k\ne i,j}^N Y^{I_k}(y_k).
\end{gather}

Because Maxwell's equation Eq.~\eqref{Maxwell1} decouples from gravity when the forms are `off-diagonal enough'---more than one index coming from a different sub-manifold---we'll never need to write down the full decomposition explicitly.  All we'll ever need are the decompositions of `diagonal' blocks and `singly off-diagonal' blocks.  They are
\begin{gather}
F_{\alpha_{1,i}\cdots\alpha_{k,i}}=\sum_I F_i^{T,I} Y_{[\alpha_{1,i}\cdots\alpha_{k,i}]}+F_i^{L,I} \nabla_{[\alpha_{1,i}} Y_{\alpha_{2,i}\cdots\alpha_{k,i}]} + F_i^{H,I} Y^H_{[\alpha_{1,i}\cdots\alpha_{k,i}]} \\
F_{\beta_j\alpha_{2,i}\cdots\alpha_{k,i}} =\sum_I F_{ji\cdots i}^{TT,I} Y_{[\alpha_{2,i}\cdots\alpha_{k,i}]\beta_j} + F_{ji\cdots i}^{TL,I} \nabla_{[\alpha_{2,i}} Y_{\alpha_{3,i}\cdots\alpha_{k,i}]\beta_j} + F_{ji\cdots i}^{TH,I} Y^H_{[\alpha_{2,i}\cdots\alpha_{k,i}]\beta_j} \hspace{.5in} \nonumber\\
\hspace{.8in}+ F_{ji\cdots i}^{LT,I} \nabla_{\beta_j}Y_{[\alpha_{2,i}\cdots\alpha_{k,i}]} + F_{ji\cdots i}^{LL,I} \nabla_{\beta_j}\nabla_{[\alpha_{2,i}} Y_{\alpha_{3,i}\cdots\alpha_{k,i}]} + F_{ji\cdots i}^{LH,I} \nabla_{\beta_j}Y^H_{[\alpha_{2,i}\cdots\alpha_{k,i}]},
\end{gather}
where we have used the fact that there are no harmonic 1-forms on a positive curvature manifold.

Also as before, we will be able to take the components that are longitudinal on the sub-manifolds and construct combinations that are transverse on the whole manifold.  To see this, take the divergence:
\begin{gather}
\sum_j\nabla^{\beta_j}F_{\beta_j\alpha_{2,i}\cdots\alpha_{k,i}}=\sum_I \left(\sum_{j\ne i}\lambda_j F_j^{LT,I}+\frac1{k-1}\tau^{(k-1)} F_i^{L,I}\right)Y_{[\alpha_{2,i}\cdots\alpha_{k,i}]} \nonumber\\
+ \left(\sum_{j\ne i}\lambda_j F_j^{LL,I} \right) \nabla_{[\alpha_{2,i}} Y_{\alpha_{3,i}\cdots\alpha_{k,i}]} +  \left(\sum_{j\ne i}\lambda_j F_j^{LH,I} \right)Y^H_{[\alpha_{2,i}\cdots\alpha_{k,i}]}.
\end{gather}

\noindent Because the basis is orthonormal, enforcing transversality on the whole manifold enforces three distinct conditions for each $I$:
\begin{gather}
\sum_{j\ne i}\lambda_j F_j^{LT,I}=-\frac1{k-1}\tau^{(k-1)} F_i^{L,I}, \nonumber\\ \sum_{j\ne i}\lambda_j F_j^{LL,I}  =0, \hspace{.3in} \text{and}\hspace{.3in}  \sum_{j\ne i}\lambda_j F_j^{LH,I}=0.
\end{gather}

To be harmonic on the whole manifold, each term in the product must be harmonic (zero-mode scalars count as harmonic).  This confirms the K\"{u}nneth formula, which tells us the $k$th Betti number $b_k(Z)$ of a product manifold $Z=Z_1\times\cdots \times Z_N$
\begin{gather}
b_k(Z)=\sum_{\substack{k_1\cdots k_N,  \\ \text{with} \;\; k_1+\cdots+ k_N=k}} b_{k_1}(Z_1)\;\cdots\; b_{k_N}(Z_N).
\end{gather}

\subsection{Decomposing the Fluctuations}
\label{decomp}

We can now give the explicit decompositions of our fluctuation fields $h_{MN}$ and $B_q$.

\textbf{Decomposition of the gravity fluctuations:}
For $h_{MN}$  it will be helpful to pull off the $N$ sub-traces\footnote{Let's slip in a quick reminder here that parentheses not only symmetrize, but they also trace-subtract.} which we define as $h_i$.
\begin{equation}
h_{(\alpha_i\beta_i)}=h_{\alpha_i\beta_i}-\frac1qh_i \;g_{\alpha_i\beta_i}.
\end{equation}
The $h_i$ can be thought of as controlling the radius and shape of the $M_{q,i}$.
It will also be helpful to shift the metric fluctuations for the $p$ extended dimensions by defining
\begin{equation}
\label{Weylshift}
H_{\mu\nu}\equiv h_{\mu\nu}+\frac1{p-2}\sum_i h_i g_{\mu\nu}.
\end{equation}
This shift is the linearized version of the Weyl transform that takes you to Einstein frame, and it will cancel the contributions of the extra-dimensional curvature to the $(\mu\nu)$ component of Einstein's equation.  Because we make this shift, most of our results don't necessarily apply to the $p=2$ case, where Einstein frame is not available; only equations without any factors of $H_{\mu\nu}$ remain valid when $p=2$.  Of course, we will pull off the trace\footnote{Reminding you again here might be overkill.}
\begin{equation}
H_{(\mu\nu)}=H_{\mu\nu}-\frac1pH g_{\mu\nu}.
\end{equation}

We are now ready to present the decomposition of the linearized fluctuations:
\begin{eqnarray}
H_{(\mu\nu)}(x,y)&=&\sum_I H_{(\mu\nu)}^I(x) Y^I(y) \nonumber \\
H(x,y)&=&\sum_I H^I(x) Y^I(y) \nonumber \\
h_i(x,y)&=&\sum_I h_i^I(x) Y^I(y) \nonumber \\
h_{\mu\alpha_i}(x,y)&=&\sum_I C_{\mu,i}^{T,I}(x) Y^I_{\alpha_i}(y) +C_{\mu,i}^{L,I}(x) \nabla_{\alpha_i}Y^I(y) \nonumber \\
h_{(\alpha_i\beta_i)}(x,y)&=&\sum_I \phi_i^{TT,I}(x) Y^I_{(\alpha_i\beta_i)}(y) + \phi_i^{TL,I}(x) \nabla_{(\alpha_i}Y^I_{\beta_i)}(y)+ \phi_i^{LL,I}(x) \nabla_{(\alpha_i}\nabla_{\beta_i)}Y^I(y) \nonumber \\
h_{\alpha_i\beta_j}(x,y)&=&\sum_I \theta_{ij}^{TT,I}(x) Y^I_{\alpha_i\beta_j}(y) + \theta_{ij}^{LT,I}(x) \nabla_{\alpha_i}Y^I_{\beta_j}(y)+\theta_{ij}^{TL,I}(x) \nabla^{}_{\beta_j}Y^I_{\alpha_i}(y)\nonumber \\
&\;& \hspace{5cm} + \theta_{ij}^{LL,I}(x) \nabla_{\alpha_i}\nabla_{\beta_j}Y^I(y).
\label{decomph}
\end{eqnarray}

\newpage

Components with both indices on the internal manifold are decomposed as a symmetric 2-tensor, with $\phi_i$ referring to diagonal blocks, and $\theta_{ij}$ referring to off-diagonal blocks.  Components with a single index on the internal manifold are decomposed like a vector, which we call $C_\mu$.  Finally, components with both indices along the extended dimensions are decomposed as scalars on the internal manifold.  When the internal manifold is a product of $N$ $q$-spheres, the eigenvector is indexed by a list of $N$ angular momenta $I=(\ell_1,\;\dots\;,\;\ell_N)$.  
For example, if $I=(\ell_1=0,\;\ell_2=5)$, $h_1^I$ corresponds to varying the size of the first sphere as you move around the second, and $h_2^I$ corresponds to changing the shape of the second sphere uniformly in the first.

Symmetry of the metric enforces
\begin{equation}
\theta_{ij}^{TT,I}=\theta_{ji}^{TT,I}, \hspace{1.3cm}  \theta_{ij}^{LT,I}=\theta_{ji}^{TL,I}, \hspace{1.3cm} \theta_{ij}^{LL,I}=\theta_{ji}^{LL,I}.
\end{equation}
(Most) of the gauge freedom can be fixed by enforcing transversality part that remains after the sub-traces have been subtracted off:
\begin{equation}
\sum_{i=1}^N\nabla^{\alpha_i} h_{\alpha_i\mu}=0 \hspace{.5in} \text{and}\hspace{.5in} \sum_{i=1}^N\nabla^{\alpha_i} h_{\alpha_i\beta_j}=\frac1q\nabla_{\beta_j} h_j,
\end{equation}
for a total of $D$ gauge-fixing conditions.

This gauge choice enforces
\begin{gather}
\sum_{i=1}^N \lambda_i C_{\mu,i}^{L,I}=0,
\label{condition1} \\ 
\sum_{i=1, i\ne j}^N \lambda_i \theta_{ij}^{LT,I}=-\left(\frac12\kappa_j+\frac{q-1}{R_j^{\;2}}\right) \phi_j^{TL,I}, 
\label{condition2}\\
\sum_{i=1, i\ne j}^N \lambda_i \theta_{ij}^{LL,I}=-\left(\frac{q-1}q\lambda_j+\frac{q-1}{R_j^{\;2}}\right) \phi_j^{LL,I}.
\label{condition3}
\end{gather}
If $Y_{\alpha_j}^{I=K}$ is a Killing vector, whose eigenvalue $\kappa_j^{I=K}=-2(q-1)/R_j^{\;2}$ saturates the Lichnerowicz bound,  then the right-hand side of Eq.~\eqref{condition2} is zero; Killing vectors satisfy Eq.~\eqref{KillingVec}, and therefore $\phi_j^{TL,I=K}$ doesn't contribute to $h_{MN}$.  Likewise, if $Y^{I=C}$ is a conformal scalar, whose eigenvalue $\lambda_j^{I=C}=-q/R_j^{\;2}$ saturates the Lichnerowicz bound on excited modes, then the right-hand side of Eq.~\eqref{condition3} is zero; conformal scalars satisfy Eq.~\eqref{ConformalScalar}, and therefore $\phi_j^{LL,I}$ doesn't contribute to $h_{MN}$.

\newpage

How much gauge freedom is left?  Under a linearized gauge transformation, $\delta h_{MN}=\nabla_M\xi_N(x,y)+\nabla_N\xi_M(x,y)$, so a diffeomorphism not fixed by our gauge choice must satisfy
\begin{gather}
\delta \left[\sum_{i=1}^N\nabla^{\alpha_i}h_{\alpha_i\beta_j}-\frac1q\nabla_{\beta_j}h_j\right]=0 \implies  \sum_i 2\nabla^{\alpha_i}\nabla_{(\alpha_i}\xi_{\beta_j)}(x,y)=0,
\\
\delta \left[\sum_{i=1}^N\nabla^{\alpha_i}h_{\mu\alpha_i}\right]=0 \implies \Box_y\xi_\mu(x,y)+\nabla_\mu\sum_{i=1}^N\nabla^{\alpha_i}\xi_{\alpha_i}(x,y)=0.
\end{gather}
The first equation is solved by $\xi_\alpha(x,y)=\phi (x) \times V_\alpha(y)$, where $V_\alpha(y)$ is either a constant, a Killing vector $Y_\alpha^{I=K}$, or the gradient of a conformal scalar $\nabla_\alpha Y^{I=C}$.  If $V_\alpha$ is either a constant or a Killing vector, then the second equation is solved when $\xi_\mu(x,y)=\psi_\mu(x) Y^{I=0}(y)$; if $V_\alpha$ is the divergence of a conformal scalar, then the second equation is solved by $\xi_\mu(x,y)=\nabla_\mu\phi(x) Y^{I=C}(y)$.

This means that there's residual, unfixed gauge invariance for the zero-mode sector, the Killing-vector sector, and the conformal-scalar sector.  For the zero-mode sector, this extra gauge invariance shifts $\delta h_{\mu\nu}^{I=0}=(\nabla_\mu\psi_\nu+\nabla_\nu\psi_\mu) Y^{I=0}$;  we find linearized diffeomorphism invariance for $h_{\mu\nu}^{I=0}$, which allows for the existence of a massless $p$-dimensional graviton.  For the Killing-vector sector, this extra gauge invariance shifts $\delta C_\mu^{I=K}=\nabla_\mu \phi(x) Y^{I=K}$; we find linearized U(1) gauge invariance, which allows for the existence of a massless vector field for every Killing vector.  Finally, for the conformal-scalar sector, the extra gauge invariance shifts $\delta h_i^{I=C}=2\lambda^{I=C} \phi(x) Y^{I=C}$.  This is our first indication that conformal scalars will require special treatment.  We will see below that, for the conformal-scalar sector, we will be able to construct a single gauge invariant combination of gravity and form field fluctuations.

\textbf{Decomposition of the form field fluctuations:}
Maxwell's equation only couple to gravity when all but one index is from the same sub-manifold, and we'll therefore only need explicit decompositions for those modes:
\begin{eqnarray}
B_{\alpha_{2,i}\cdots\alpha_{q,i}}(x,y) &=&\sum_I b_i^{T,I}(x) Y^I_{[\alpha_{2,i}\cdots\alpha_{q,i}]}(y) + b_{i}^{L,I}(x) \nabla_{[\alpha_{2,i}}Y^I_{\alpha_{3,i}\cdots\alpha_{q_i}]}(y), \nonumber \\
B_{\beta_j \alpha_{3,i}\cdots\alpha_{q,i}} (x,y) &=&\sum_I \beta_{ji\cdots i}^{TT,I}(x) Y^I_{[\alpha_{3,i}\cdots\alpha_{q,i}]\beta_j}(y) +\beta_{ji\cdots i}^{TL,I}(x) \nabla_{[\alpha_{2,i}}Y^I_{\alpha_{3,i}\cdots\alpha_{q,i}]\beta_j}(y) \nonumber \\ &&\;\;\;\; + \; \beta_{ji\cdots i}^{TH,I}(x) Y^{H,I}_{[\alpha_{3,i}\cdots\alpha_{q,i}]}(y) 
 + \beta_{ji\cdots i}^{LT,I}(x) \nabla_{\beta_j}Y^I_{[\alpha_{2,i}\cdots\alpha_{q,i}]}(y)  
 \nonumber\\&&\;\; \;\;
 +\; \beta_{ji\cdots i }^{LL,I}(x)\nabla_{\beta_j}\nabla_{[\alpha_{2,i}}Y^I_{\alpha_{3,i}\cdots\alpha_{q_i}]}(y) +\beta_{ji\cdots i}^{LH,I}(x) \nabla_{\beta_j}Y^{H,I}_{[\alpha_{2,i}\cdots\alpha_{q,i}]}(y), \nonumber \\
B_{\mu\alpha_{1,i}\cdots\alpha_{q-2,i}} (x,y)&=&\sum_I b_{\mu,i}^{T,I}(x) Y^I_{[\alpha_{1,i}\cdots\alpha_{q-2,i}]}(y)+b_{i}^{L,I}(x) \nabla_{[\alpha_{1,i}}Y^I_{\alpha_{2,i}\cdots\alpha_{q-2,i}]}(y)\nonumber\\ &&\;\;\;\; +\; b_{\mu,i}^{H,I}(x) Y^{H,I}_{[\alpha_{1,i}\cdots\alpha_{q-2,i}]}(y).
\label{decompB}
\end{eqnarray}
The components with no indices on $M_p$ are decomposed as $(q-1)$-forms; we've used the notation $b_i$ for diagonal blocks and $\beta_{ji\cdots i}$ for singly off-diagonal blocks.  The components with one index on $M_p$ are decomposed as $(q-2)$-forms; they are vectors from the perspective of the $M_p$.  

(Most) of the gauge freedom can be fixed by enforcing transversality:
\begin{gather}
\sum_{k=1}^N\nabla^{\beta_k}B_{\beta_k\alpha_{2,j}\cdots\alpha_{q-1,j}}=0, \hspace{.3in} \sum_{k=1}^N\nabla^{\beta_k}B_{\beta_k\mu\alpha_{3,j}\cdots\alpha_{q-1,j}}=0, \hspace{.3in} \\
\text{and, more generally,} \hspace{.3in} d^{\dagger}_{Nq} B=0.
\end{gather}
For a total of $D^{q-2}$ gauge fixing conditions.  These conditions enforce
\begin{gather}
\label{condition4}
\sum_{k=1,k\ne j}^N \lambda_k\beta_{ji\cdots i}^{LT,I}=-\frac1{q-1}\kappa_j b_j^{L,I} \\
\label{condition5}
\sum_{k=1,k\ne j}^N \lambda_k\beta_{ji\cdots i}^{LL,I}=0 \\
\label{condition6}
\sum_{k=1,k\ne j}^N \lambda_k\beta_{ji\cdots i}^{LH,I}=0.
\end{gather}

How much gauge invariance is left? Our gauge choice doesn't touch the sector proportional to harmonic forms; there is therefore enough residual gauge invariance to account for a massless excitation for every harmonic $k$-form.  (There is a trivial constant harmonic $q$-form on each $q$-dimensional $M_{q,i}$, and the $b_i^{T,I=0}$ therefore have residual gauge invariance.  However, because the equations of motion depend only on $f=dB$, the zero-mode fluctuation $b_i^{T,I=0}$ never appears; there is no physical zero-mode fluctuation of $f$ because such a mode would change the number of flux lines, which is a conserved quantity.)

Finally, we can return to the question of the extra gauge invariance in the conformal-scalar sector.  Notice that $F_{M_1\cdots M_q}$ shifts under linearized diffeomorphisms, and in particular that under residual conformal-scalar gauge invariance, $F_{\alpha_{1,i}\cdots\alpha_{q,i}}\epsilon^{\alpha_{1,i}\cdots\alpha_{q,i}}\rightarrow c_i+b_i^{T,I=C}\phi(x)Y^{I=C}$.  Perturbing a sphere by its $\ell=1$ mode shifts the sphere to the left, leaving the metric invariant; if the flux moves with it, that mode is gauge.  On the other hand, if the sphere shifts left while the flux sloshes right, that is a physical perturbation of the background.  The combination $h_i^{I=C}-2\lambda^{I=C}b_i^{T,I=C}/c_i$ is the one gauge invariant combination of $h_i^{I=C}$ and $b_i^{T,I=C}$; it corresponds to sloshing the flux in the opposite direction to the shift of the sphere.

\subsection{Equations of Motion for the Fluctuations}
\label{eom1}
Now we can get to work.  In this subsection, we plug the decompositions given above---Eqs.~\eqref{decomph} and \eqref{decompB}---into the first-order equations of motion---Eqs.~\eqref{Einstein1} and \eqref{Maxwell1}.  Because the Lichnerowicz eigenbasis is orthonormal, the components of the equations that lie along each eigenvector must be true separately.  
We will use $\Box_x=\nabla^\mu\nabla_\mu$ as the Laplacian on the $p$ extended dimensions and $\triangle_y$ as the Lichnerowicz operator on the $Nq$ internal dimensions; we'll also use $\triangle_k$ as the Lichnerowicz operator restricted to the $k$th sub-manifold, $M_{q,k}$.  We also define $\text{Max} \;T_\mu\equiv\Box_x T_\mu -\nabla^\rho\nabla_\mu T_\rho$ as the Maxwell operator acting on vectors on $M_p$---acting on a divergence-free vector, it could equally well be written as $\triangle_x$.

 From the $(\mu,\nu)$ sector of Einstein's equation, we get
\begin{equation}
\label{E11}
\left[R_{\mu\nu}^{\;\;(1)}(H_{\rho\sigma}^I)-\frac12\Box_y H_{\mu\nu}^I-\frac{p-1}{L^2}H_{\mu\nu}^I+A g_{\mu\nu}\right] Y^I=0,
\end{equation}
where $R_{\mu\nu}^{\;\;(1)}(H_{\rho\sigma}^I)$ is the linearized Ricci tensor for only the extended $p$ dimensions, and we've defined the quantity $A$ as
\begin{gather}
A\equiv \sum_{k=1}^N \left[\frac12\frac1{p-2}\left(\Box_x+\Box_y + 2\frac{p-1}{L^2}\right) h_k^I 
-\frac12\frac{q-1}{D-2} c_k^{\;2}\left(h_k^I-\frac2{c_k}\Box_kb_k^{T,I}\right)\right].
\label{E11A}
\end{gather}

From the $(\mu,\alpha_i)$ sector of Einstein's equation, we get
\begin{equation}
\label{E21}
\left[\left(\text{Max} +\triangle_y+2\frac{p-1}{L^2}\right)C_{\mu,i}^{T,I}+\triangle_i\left(c_ib_{\mu,i}^{T,I}-\frac1{q-1}\nabla_\mu c_i b_i^{L,I}\right)\right]Y_{\alpha_i}^I=0,
\end{equation}
and
\begin{gather}
 \Bigg[\left(\text{Max} +\triangle_y+2\frac{p-1}{L^2}\right)C_{\mu,i}^{L,I}-\nabla^\rho H_{\rho\mu}^I 
+\nabla_\mu \left(H^I -\frac1{p-2}\sum_{k=1}^N h_k^I  -\frac1q h_i^I+c_ib_i^{T,I} \right) \Bigg]\nabla_{\alpha_i}Y^I=0.
\label{E22}
\end{gather}

From the $(\alpha_i, \beta_i)$ sector of Einstein's equation, we get
\begin{gather}
\label{E31}
\left[\left(\Box_x+\triangle_y+2\frac{q-1}{R_i^{\;2}}\right)\phi_i^{TT,I}\right]Y_{(\alpha_i\beta_i)}^I=0, \\
\label{E32}
\left[\left(\Box_x+\triangle_y+2\frac{q-1}{R_i^{\;2}}\right)\phi_i^{TL,I} - 2\nabla^\mu C_{\mu,i}^{T,I}\right]\nabla_{(\alpha_i}Y_{\beta_i)}^I=0, \\
\left[\left(\Box_x+\Box_y+2\frac{q-1}{R_i^{\;2}}\right)\phi_i^{LL,I} +\left(H^I-\frac{2}{p-2}\sum_{k=1}^Nh_k^I -\frac2qh_i^I- 2\nabla^\mu C_{\mu,i}^{L,I}\right)\right]\nabla_{(\alpha_i}\hspace{-.5mm} \nabla_{\beta_i)}Y^I=0,
\label{E33}
\end{gather}
and
\begin{gather}
\Bigg[\left(\Box_x+\Box_y+2\frac{q-1}{R_i^{\;2}}\right)h_i^I+\Box_i\left(H^I-\frac{2}{p-2}\sum_{k=1}^Nh_k^I-\frac2qh_i^I-2\nabla^\mu C_{\mu,i}^{L,I}\right) \nonumber\\
-q\left(c_i^{\;2}h_i^I-2\Box_ic_ib_i^{T,I}\right)+q\frac{q-1}{D-2}\sum_{k=1}^N \left(c_k^{\;2}h_k^I-2\Box_kc_kb_k^{T,I}\right)\Bigg]g_{\alpha_i\beta_i}Y^I=0.
\label{E34}
\end{gather}

Finally, from the $(\alpha_i, \beta_j)$, with $i\ne j$, sector of Einstein's equation, we get 
\begin{gather}
\label{E41}
\left[\left(\Box_x+\triangle_y +2\frac{p-1}{L^2}\right)\theta_{ij}^{TT,I}+\triangle_j c_j \beta_{ij\cdots j}^{TT,I} + \triangle_i c_i \beta_{ji\cdots i}^{TT,I}\right]Y_{\alpha_i\beta_j}^I =0,\\
\label{E42}
\left[\left(\Box_x+\triangle_y +2\frac{p-1}{L^2}\right)\theta_{ij}^{LT,I}-\nabla^\mu C_{\mu,j}^{T,I}+\triangle_j \left(c_j \beta_{ij\cdots j}^{LT,I}-\frac{1}{q-1}c_j b_j^{L,I}\right)\right]\nabla_{\alpha_i}Y_{\beta_j}^I =0,
\end{gather}
and
\begin{gather}
\Bigg[\left(\Box_x+\triangle_y +2\frac{p-1}{L^2}\right)\theta_{ij}^{LL,I}+\frac12\left(H^I-\frac2{p-2}\sum_{k=1}^Nh_k^I-\frac2qh_i^I -2\nabla^\mu C_{\mu,i}^{L,I} \right) \nonumber\\
+\frac12\left(H^I-\frac2{p-2}\sum_{k=1}^Nh_k^I-\frac2qh_j^I-2\nabla^\mu C_{\mu,j}^{L,I}\right)+c_ib_i^{T,I}+c_jb_j^{T,I}\Bigg]\nabla_{\alpha_i}\nabla_{\beta_j}Y^I=0.
\label{E43}
\end{gather}

Of Maxwell's equation, we only require the diagonal and singly off-diagonal components.  From the $(\beta_{2,i},\dots,\beta_{q,i})$ sector (after contracting with $\epsilon_{\alpha_i}^{\;\;\beta_{2,i}\cdots\beta_{q,i}}$) we get
\begin{gather}
\Bigg[ \left(\Box_x+\triangle_y\right) c_i b_i^{T,I}+\frac{c_i^{\;2}}{2}\left(H-\frac2{p-2}\sum_{k=1}^N h_k^I-\frac2qh_i-2\nabla^\mu C_{\mu,i}^{L,I}\right)
 -\frac{q-1}{q} c_i^{\;2} h_i^{I} 
 \nonumber\\
 +c_i^{\;2} \left(\frac{q-1}q\triangle_i +\frac{q-1}{R_i^{\;2}}\right)\phi_i^{LL,I}\Bigg]\nabla_{\alpha_i}Y^I=0,
 \label{M11}
\end{gather}
and
\begin{gather}
\Bigg[\triangle_i\nabla^{\mu}\left(c_ib_{\mu,i}^{T,I}-\frac1{q-1}\nabla_\mu c_ib_i^{L,I}\right)-\frac{\triangle_i}{q-1}\triangle_y c_ib_i^{L,I}-c_i^{\;2}\nabla^\mu C_{\mu,i}^{T,I} \nonumber \\
 + c_i^{\;2}\left(\frac12\triangle_i +\frac{q-1}{R_i^{\;2}}\right)\phi_i^{TL,I}\Bigg]Y_{\alpha_i}^I=0.
 \label{M12}
\end{gather}

From the $(\mu,\gamma_{3,i},\dots,\gamma_{q,i})$ sector (after contracting with $\epsilon_{\alpha_i\beta_i}^{\;\;\;\;\;\gamma_{3,i}\cdots\gamma_{q,i}}$)  we get
\begin{gather}
\label{M21}
\Bigg[\left(\text{Max}+\triangle_y\right)\left(c_i b_{\mu,i}^{T,I}-\frac1{q-1} \nabla_\mu c_ib_i^{L,I}\right)-c_i^{\;2}C_{\mu,i}^{T,I}\Bigg]\nabla_{[\alpha_i}Y_{\beta_i]}^I=0,  \\
\label{M22}
\Bigg[\left(\text{Max}+\triangle_y\right)c_i b_{\mu,i}^{L,I}-(q-2)\nabla^\nu c_i b_{\mu\nu}^{T,I}\Bigg]Y_{[\alpha_i\beta_i]}^I =0, 
\end{gather}
and
\begin{gather}
\label{M23}
\Bigg[\left(\text{Max}+\triangle_y\right)b_{\mu,i}^{H,I}\Bigg]Y_{[\alpha_i\beta_i]}^{H,I} =0.
\end{gather}

Finally, from the $(\alpha_i,\gamma_{3,j},\dots,\gamma_{q,j})$ sector (after contracting with $\epsilon_{\beta_j\delta_j}^{\;\;\;\;\;\gamma_{3,i}\cdots\gamma_{q,i}}$) we get
\begin{gather}
\label{M31}
\Bigg[\left(\Box_x+\triangle_y\right)c_j\beta_{ij\cdots j}^{TT,I}-c_j^{\;2}\theta_{ij}^{TT,I}\Bigg]\nabla_{[\beta_j}Y_{\delta_j]\alpha_i} =0,\\
\label{M32}
\Bigg[\left(\Box_x+\triangle_y\right)c_j\beta_{ij\cdots j}^{LT,I}-c_j\nabla^\mu b_{\mu,j}^{T,I}-c_j^{\;2}\theta_{ij}^{LT,I}\Bigg]\nabla_{\alpha_i}\nabla_{[\beta_j}Y_{\delta_j]} =0,\\
\label{M33}
\Bigg[(\Box_x+\triangle_y)\beta_{ij\cdots j}^{TL,I}-(q-2)\nabla^\mu\beta_{\mu,ij\cdots j}^{TT,I}\Bigg]Y^I_{[\beta_j\delta_j]\alpha_i}=0,\\
\label{M34}
\Bigg[(\Box_x+\triangle_y)\beta_{ij\cdots j}^{LL,I}-(q-2)\nabla^\mu\left(\beta_{\mu,ij\cdots j}^{LT,I}-\frac1{q-2}b_{\mu,j}^{L,I}\right)\Bigg]\nabla_{\alpha_i}Y_{[\beta_j\delta_j]}^I=0,\\
\label{M35}
\Bigg[(\Box_x+\triangle_y)\beta_{ij\cdots j}^{TH,I}\Bigg]Y_{[\beta_j\delta_j]\alpha_i}^{H,I}=0,
\end{gather}
and
\begin{gather}
\label{M36}
\Bigg[(\Box_x+\triangle_y)\beta_{ij\cdots j}^{LH,I}-\nabla^\mu b_{\mu,j}^{H,I}\Bigg]\nabla_{\alpha_i}Y_{[\beta_j\delta_j]}^{H,I}=0.
\end{gather}

For the components that are more off-diagonal, the equation of motion can be written as
\begin{gather}
d^\dagger d B_{q-1}=\left(\triangle_x+\triangle_y-dd_p^\dagger\right)B_{q-1}=0,
\label{M41}
\end{gather}
where we've exchanged the $d$ and $d^\dagger$ and used the gauge-fixing condition $d^{\dagger}_{Nq} B_{q-1}=0$.  Equations~\eqref{M23} and \eqref{M33}-\eqref{M36} are of precisely this form because they are decoupled from metric fluctuations.

Our two partial differential equations have been broken up into many ordinary differential equations---bite-sized pieces we will devour in Sec.~\ref{sec:diagonalizing}.

\newpage

\section{The Spectrum and Stability}
\label{sec:diagonalizing}  

Equations~\eqref{E11}-\eqref{M41} are now ordinary, coupled, second-order differential equations for the fluctuations.  The last step in finding the spectrum of our compactification will be to diagonalize, which brings the equations into the appropriate form for massive scalars, vectors, gravitons, and $k$-forms---
\begin{gather}
\Box_x \phi(x)=m^2\phi(x), \nonumber\\
\text{Max} \;V_\mu(x)=m^2 V_\mu(x) \hspace{.3in} \text{and}\hspace{.3in} \nabla^\mu V_\mu=0, \nonumber \\
\Box_x T_{\mu\nu}(x)=\left(m^2-\frac2{L^2}\right)T_{\mu\nu}(x), \hspace{.3in} \nabla^\mu T_{\mu\nu}=0 \hspace{.3in} \text{and}\hspace{.3in} T^\mu_{\;\;\mu}=0, \nonumber\\
\triangle_x F_k = m^2 F_k\hspace{.3in} \text{and}\hspace{.3in} d^{\dagger}_p F_k=0
\end{gather}
---from which we can read off the masses.

In Sec.~\ref{0modes}, we will first look at the zero-mode sector, where we will find a massless graviton and the equation of motion for the radii of the $N$ internal manifolds.  This latter equation is where the `total-volume instability' appears; we will show that as in the $N=1$ case, there is a  range of vacua for which this mode is explicitly stabilized.  In Sec.~\ref{coupleddiagonalscalars}, we will look at the coupling of the diagonal scalars $h_i$ and the $b_i^T$.  It's here that the `lumpiness instability' appears; we will show that as in the $N=1$ case, compactifications involving 2- and 3-spheres are always stable, but compactifications involving higher spheres can have instabilities.  In Sec.~\ref{offdiagonalscalars}, we will look at the scalars that come from off-diagonal components---the $\phi_i$, the $\theta_{ij}$ and the $\beta_{ij\cdots j}$ modes.  This is the sector where the `cycle-collapse instability' appears. In Sec.~\ref{oneforms}, we discuss the coupling of the graviphoton $C_\mu$ to the one-form fluctuations $b_\mu$.  We find a massless vector for each Killing vector of the internal manifold, plus extra massless vectors for each harmonic 2-form; the other vector fluctuations are all massive.  In Sec.~\ref{massivegraviton}, we find massive tensor fluctuations whose mass is always above the Higuchi bound for consistent propagation of a massive graviton \cite{Higuchi:1986py}.  Finally, in Sec.~\ref{higherforms}, we find the masses of the remaining form fluctuations, which are decoupled from gravity.  Threats to stability arise only in Sec.~\ref{0modes} and Sec.~\ref{coupleddiagonalscalars}.

\subsection{ The Zero-Mode Sector} 
\label{0modes}
Let's first look at the zero modes, the equations of motion proportional to $Y^{I=0}$.  These are Eqs.~\eqref{E11} and~\eqref{E34} with all the $y$-derivatives set to zero, which can be written:
\begin{equation}
R_{\mu\nu}^{(1)}(H_{\rho\sigma}^{I=0})-\frac{p-1}{L^2}H_{\mu\nu}^{I=0}=0,
\label{masslessgraviton}
\end{equation}
and 
\begin{equation}
\label{radiizeromodes}
\Box_x h_i^{I=0}=\left(qc_i^{\;2}-2\frac{q-1}{R_i^{\;2}}\right)h_i^{I=0}-q\frac{q-1}{D-2}\sum_{k=1}^N c_k^{\;2}h_k^{I=0},
\end{equation}
where $R_{\mu\nu}^{(1)}(H_{\rho\sigma})$ is the linearized Ricci tensor just for the $p$ extended dimensions, and the superscript $I=0$ denotes the zero mode.

Eq.~\eqref{masslessgraviton} is the equation for a massless spin-2 particle propagating on a maximally symmetric spacetime with radius of curvature $L$.  The transverse and traceless part of Eq.~\eqref{masslessgraviton} can be written \begin{gather}
\Box_x H_{(\mu\nu)}^{I=0}=-\frac2{L^2} H_{(\mu\nu)}^{I=0}.
\end{gather}
A massless graviton propagating on a curved background has an apparent mass of $-2/L^2$ \cite{vanNieuwenhuizen:1984iz,Hinterbichler:2011tt}.  Together with the surviving diffeomorphism invariance for the zero mode demonstrated in Sec.~\ref{decomp}, we find the expected $p$-dimensional massless graviton.

Eq.~\eqref{radiizeromodes} is the equation of motion for small fluctuations in the radii of the $N$ sub-manifolds.  It's here that the `total-volume instability' is found.  Before discussing the case of general $N$, let's return briefly to $N=1$.

\subsubsection{The N=1 Zero-Mode Sector}
We can solve the background equations of motion Eqs.~\eqref{0sol1} and \eqref{0sol2} to get:
\begin{gather}
\label{cR}
c^2 R^2=\frac{2(D-2)(q-1)}{p-1}-\frac{R^2}{p-1}(4\Lambda).
\end{gather}
In Fig.~4, $cR$ is plotted as a function of $R$.  When $\Lambda=0$, $cR$ is a constant;  when $\Lambda<0$, $cR$ grows with $R$; and when $\Lambda>0$, $cR$ falls and hits 0 at a finite value of $R$.  This solution with $cR=0$ and $R>0$ is the uncharged Nariai solution.  Independent of $\Lambda$, when $R$ goes to zero, $cR$ approaches the same constant.  This solution, which we dub the `nothing state' in \cite{Brown:2011gt}, is so overwhelmed by curvature and flux density that the effects of nonzero $\Lambda$ are inconsequential.
The background equations of motion also tell us that $M_p$ is de Sitter when
\begin{gather}
\label{dSvacuum}
\text{de Sitter:} \hspace{.3in} c^2R^2< 2(q-1).
\end{gather}



Now let's consider zero-mode fluctuations about this background.  There is a single fluctuation $h^{I=0}$, which can be identified with fluctuations in the total volume of the internal manifold.  Its equation of motion is
\begin{gather}
\Box_x h^{I=0}=\frac1{R^2}\left(-2(q-1)+ \frac{q(p-1)}{D-2} c^2R^2\right)h^{I=0}.
\end{gather}
The zero-mode therefore has a positive mass whenever
\begin{gather}
\label{N1zeromodestability}
\text{zero-mode stability:} \hspace{.3in} c^2R^2> 2\frac{q-1}{q}\frac{D-2}{p-1}.
\end{gather}

Comparing this zero-mode stability condition Eq.~\eqref{N1zeromodestability} against the de Sitter condition in Eq.~\eqref{dSvacuum} shows that all AdS and Minkowski vacua are stable against zero-mode fluctuations---indeed these fluctuations always have positive mass.  When $\Lambda>0$, some de Sitter solutions are stable and some are unstable.  The stable de Sitter solutions correspond to the small-volume minima of the effective potential in Fig.~1, and unstable de Sitter solutions correspond to the large-volume maxima. The zero-mode instability of the Nariai solution is precisely the negative mode identified in \cite{Ginsparg:1982rs}  that leads to the nucleation of extremal black ($p-2)$-branes in de Sitter space.  
%
%
%
%
%
%

\begin{figure}[t] 
   \centering
   \includegraphics[width=1 \textwidth]{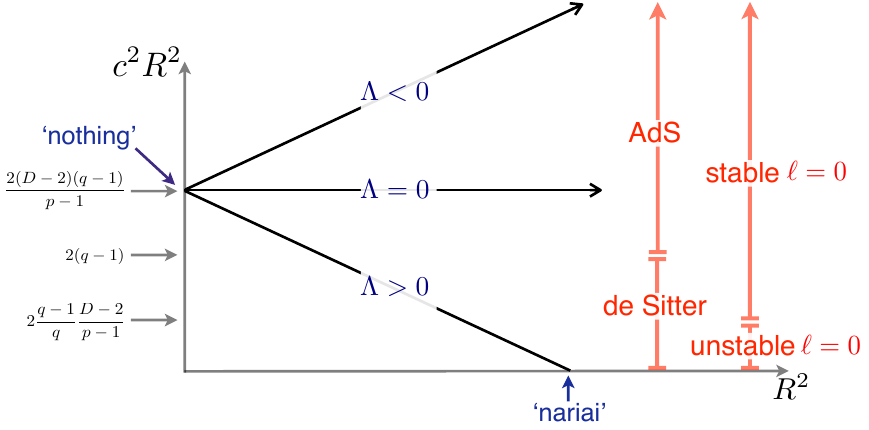} \label{fig:fig2}
   \caption{A plot of $c^2R^2$ vs. $R^2$ for $N=1$.  When $\Lambda=0$, $c^2R^2$ is constant; when $\Lambda<0$, $c^2R^2$ is a growing function of $R^2$; when $\Lambda>0$, $c^2R^2$ falls and hits zero at a finite value of $R^2$.  For any $\Lambda$, the state with $R=0$ is called the `nothing state' and has the same value of $c^2R^2$.  The state with $c^2R^2=0$ and $R>0$ (meaning $c=0$) is called the Nariai solution.    Also plotted are the conditions for having a de Sitter compactification, and for stability against zero-mode perturbations.}
\end{figure}



\subsubsection{The $\bm{N \geq2}$ Zero-Mode Sector}
\label{0modesN}
We can solve the background equations of motion Eqs.~\eqref{0sol1} and \eqref{0sol2} to get:
\begin{gather}
\label{cRN}
c_i^{\;2} R_i^{\;2}=\frac{2(D-2)(q-1)} {D-q-1} -\frac{R_i^{\;2}}{D-q-1}\left(4\Lambda -(q-1)\sum_{k=1,k\neq i}^Nc_k^{\;2}\right).
\end{gather}
This formula is analogous to Eq.~\eqref{cR}, except for the new term which accounts for the contribution of the flux around the other sub-manifolds involved in the compactification.   Flux wrapped around the other sub-manifolds has the same impact on $c_iR_i$ as a negative contribution to the cosmological constant.  The background field equations also tell us that $M_p$ is de Sitter when
\begin{gather}
\label{dSvacuumN}
\text{de Sitter:} \hspace{.3in} c_i^{\;2}R_i^{\;2}< 2(q-1);
\end{gather}
if this equation is satisfied for any $i$, then it is necessarily satisfied for all $i$.  
Every choice of a set of $c_i$'s corresponds to a solution to the background equations of motion, but some of these solutions  have negatively curved internal manifolds with $R_i^{\;-2}<0$.  Excluding these hyperbolic solutions restricts the allowed range of $c_i$'s to
\begin{gather}
\label{Rblowup}
R_i^{\;-2}\ge0:  \hspace{.3in} c_i^{\;2} \ge -2\frac{p-1}{L^2}=\frac{q-1}{D-2}\;\sum_{k=1}^N c_k^{\;2} - \frac{4}{D-2}\Lambda.
\end{gather}

Finally, a special case of interest is when all the $c_i$ are equal, $c_i=c$, and consequently all the $R_i$ are equal, $R_i=R$; in this case the entire internal manifold is an Einstein manifold.  The solution to the background equations of motion can, in this special case, be written as:
\begin{gather}
\label{cRNdiag}
c^2 R^2\Bigl |_{c_i=c}=\frac{2(D-2)(q-1)} {p+N-2}-\frac{R^{\;2}}{p+N-2} \left(4\Lambda\right).
\end{gather}

Now let's consider zero-mode fluctuations about this background.    We can write the zero-mode equation Eq.~\eqref{radiizeromodes} out more explicitly as:
\begin{gather}
\Box_x h_i^{I=0} = \sum_{j=1}^N M_{ij} h_j^{I=0},
\end{gather}
where $M_{ij}$ is an $N\times N$ matrix given by
\begin{gather}
M_{ij} =\left[\left(\begin{array}{ccc} qc_1^{\;2}-2\frac{q-1}{R_1^{\;2}} & 0 & 0 \\ 0 & \ddots & 0 \\ 0 & 0 & qc_N^{\;2}-2\frac{q-1}{R_N^{\;2}} \end{array}\right) - q\frac{q-1}{D-2}\left(\begin{array}{ccc} c_1^{\;2} & \cdots & c_N^{\;2} \\ \vdots & \ddots & \vdots \\  c_1^{\;2} & \cdots & c_N^{\;2}\end{array}\right)\right].
\end{gather}
To check stability, we need to confirm that all of the eigenvalues of this $N\times N$ matrix are positive or, if the solution is AdS, that they are less negative than the BF bound.  We will show that the zero-mode sector of the $N>1$ case is qualitatively  similar to the $N=1$ case.  In particular, we will show:
\begin{itemize}
\item All AdS and Minkowski compactifications are stable---indeed all $N$ zero-mode fluctuations have positive mass.
\item There is a range of stable de Sitter vacua.
\end{itemize}

Our proof strategy will be as follows: we will first identify a friendly solution and demonstrate that all of the eigenvalues of $M_{ij}$ for this solution are positive.  Then, to prove that another solution with a certain set of $c_i$'s is stable, we will find a path through $c_i$ space with a  positive-definite determinant  that connects this solution to the friendly one.  This will prove that the solution is stable because if we start with all positive eigenvalues, and we keep the determinant positive as we move, then we must end with all positive eigenvalues.



\paragraph{Friendly Solution: Einstein Internal Manifold, $\bm{c_i=c}$}

Our friendly solutions are ones for which the internal manifold is an Einstein manifold---all of the $c_i$ are equal, $c_i=c$, and all of the $R_i$ are equal, $R_i=R$. We can find the eigenvectors and eigenvalues of $M_{ij}$ exactly for these solutions.  The eigenvalues are 
\begin{eqnarray}
m^2&=&\frac1{R^2}\left(-2(q-1) + \frac{q(p+N-2)}{p+Nq-2} c^2R^2\right),     \hspace{.5in} \text{with multiplicity } 1, \nonumber\\
m^2&=&\frac1{R^2}\left(-2(q-1)  + q c^2R^2\right),                                \hspace{1.47in} \text{with multiplicity } N-1.
\end{eqnarray}
The first eigenvalue corresponds to the eigenvector where all of the $h_i^{I=0}$ fluctuate in unison, $\sum h_i^{I=0}$, and other $N-1$ eigenvalues corresponds to the $N-1$ volume-preserving fluctuations, such as $h_1^{I=0}-h_2^{I=0}$, $h_1^{I=0}+h_2^{I=0}-2h_3^{I=0}$, and so on.  Of these eigenvalues, the total volume fluctuation goes unstable first, and it goes unstable precisely when Eq.~\eqref{N1zeromodestability} is satisfied.  There is a range of stable de Sitter minima with
\begin{gather}
\frac{2(q-1)}{q} \frac{p+Nq-2}{p+N-2}\; <\; c^2R^2\;<\;2(q-1).
\end{gather}

\paragraph{Paths with Positive Determinant}
\label{pathszeromode}

We are interested in paths through $c_i$ space that originate at this friendly solution and preserve positivity of the eigenvalues.  Such paths have positive definite determinants.  To find them, we will use the matrix determinant lemma, which states that if $A$ is an $N\times N$ matrix and $U$ and $V$ are $N\times1$ column vectors, then
\begin{gather}
\det(A+UV^{T})=\det(A)(1+V^{T}A^{-1}U).
\end{gather}
$M_{ij}$ is of precisely this form, with
\begin{gather}
A=\left(\begin{array}{ccc} qc_1^{\;2}-2\frac{q-1}{R_1^{\;2}} & 0 & 0 \\ 0 & \ddots & 0 \\ 0 & 0 & qc_N^{\;2}-2\frac{q-1}{R_N^{\;2}} \end{array}\right), \hspace{.3 in} U=\left(\begin{array}{c} 1\\\vdots\\1\end{array}\right),\hspace{.3in} V=-q\frac{q-1}{D-2}\left(\begin{array}{c} c_1^{\;2}\\\vdots\\c_N^{\;2}\end{array}\right).
\end{gather}
The determinant of $M_{ij}$ therefore is
\begin{gather}
\label{0modedet}
\det(M_{ij})=\sum_{i=1}^N\left\{\left(\prod_{k=1,k\neq i}^N\frac{qc_k^{\;2}R_k^{\;2}-2(q-1)}{R_k^{\;2}}\right) \frac1{NR_i^{\;2}}\left[\frac{q(p+N-2)}{p+Nq-2}c_i^{\;2}R_i^{\;2}-2(q-1)\right]\right\}.
\end{gather}
Zeroes of the determinant are catastrophes of the effective potential, where two or more extrema merge and annihilate.

A sufficient condition to prove $\det(M_{ij})>0$ is that
\begin{gather}
\label{suffcond}
c_k^{\;2}R_k^{\;2}>2\frac{q-1}{q}\frac{p+Nq-2}{p+N-2} \;\;\;\;\;\;\;\;\;\;\;\forall\;k.
\end{gather}
As long as Eq.~\eqref{suffcond} is true, all the terms in the product are positive and the term in square brackets is positive, so the sum is definitely positive.  This proves that the entire strip where, for all $i$,
\begin{gather}
\label{0modestabilityN}
\frac{2(q-1)}{q} \frac{p+Nq-2}{p+N-2}\; <\; c_i^{\;2}R_i^{\;2}\;<\;2(q-1)
\end{gather}
corresponds to de Sitter minima that are stable to zero-mode fluctuations.  Because Eq.~\eqref{suffcond} is not a necessary condition for positivity of the determinant, there are other stable dS minima as well.

\subsection{Coupled Diagonal Scalars}
\label{coupleddiagonalscalars}
Now let's look at the diagonal scalars $h_i^I$ and $b_i^{T,I}$, which also couple to $\theta_{ij}^{LL,I}$.  This is the sector where the `lumpiness instability' lives.  Notice that $H^I$ never appears dynamically, so Eq.~\eqref{E33} is a constraint that we can use to eliminate $H^I$.  However, because Eq.~\eqref{E33} is proportional to $\nabla^{}_{(\alpha_i}Y^I_{\beta_i)}$, it is automatically satisfied in the conformal-scalar sector and we cannot use it to eliminate $H^{I=C}$.  Instead, to eliminate $H^{I=C}$, we can use Eq.~\eqref{M11}.  
It will be helpful to work in terms of these combinations:
\begin{eqnarray}
\bar h_i^I &=& h_i^I - 2\lambda_i\frac{b_i^{T,I} }{c_i}\nonumber\\
\bar b_i^{T,I} &=&b_i^{T,I}-\frac12 c_i \phi_i^{LL,I}\nonumber\\
\bar\theta_{ij}^{LL,T}&=&\theta_{ij}^{LL,T}-\frac{b_i^{T,I}}{c_i} -\frac{b_j^{T,I}}{c_j} .
\end{eqnarray}
In the conformal-scalar sector $I=C$, $\bar h_i^{I=C}$ is the gauge-invariant combination we identified in Sec.~\ref{decomp}.

In terms of the barred variables, Eqs.~\eqref{E34}, \eqref{M11} and \eqref{E43} can be written
\begin{eqnarray}
\label{cscalar1}
\Box_x\bar{h}_i^I&=&\left(-\sum_{k=1}^N\lambda^I_k\right)\bar{h}_i^I +\sum_{j=1}^NM_{ij}\bar{h}_j^I -2\frac{q-1}q\lambda^I_i\bar{h}_i^I-\frac{4}{c_i^{\;2}}\frac{q-1}q\lambda^I_i\left(\lambda^I_i+\frac{q}{R_i^{\;2}}\right) c_i\bar b_i^{T,I}\\
\label{cscalar2}
\Box_x c_i \bar b_i^{T,I}&=&\left(-\sum_{k=1}^N\lambda^I_k\right)c_i\bar{b}_i^{T,I} + c_i^{\;2}\frac{q-1}{q} \bar h_i^I+2 \frac{q-1}{q}\lambda^I_i c_ib_i^{T,I} \\
\label{cscalar3}
\Box_x\bar\theta_{ij}^{LL,I}&=&\left(-\sum_{k=1}^N\lambda^I_k-2\frac{p-1}{L^2}\right)\bar\theta_{ij}^{LL,I} +\frac{q-1}{q}\left(\bar h_i^I +2\lambda_i\frac{\bar{b}_i^{T,I}}{c_i} + \bar h_j^I +2\lambda_j\frac{\bar{b}_j^{T,I}}{c_j}\right)\nonumber\\
&&\hspace{1.85in}+ 2\frac{q-1}{R_i^{\;2}} \frac{\bar{b}_i^{T,I}}{c_i}+2\frac{q-1}{R_j^{\;2}} \frac{\bar{b}_j^{T,I}}{c_j}.
\end{eqnarray}
For the first and third equations, we used Eq.~\eqref{M11} to eliminate $H^I$; only for the middle equation did we use the constraint.  Therefore: Eq.~\eqref{cscalar1} is applicable in all sectors; Eq.~\eqref{cscalar2} is not applicable in the zero-mode sector or the conformal-scalar sector, because in either case $\bar b_i$ is not a dynamical fluctuation; and Eq.~\eqref{cscalar3} is only applicable when both $\lambda_i^I$ and $\lambda_j^I$ are excited, because otherwise $\theta_{ij}^{LL,I}$ is not a physical fluctuation of $h_{\alpha\beta}$.  Notice that the last term in Eq.~\eqref{cscalar1}, which couples $\bar h_i^I$ to $\bar b_i^{T,I}$, goes to zero for either zero modes (with $\lambda_i^{I=0}=0$) or conformal scalars (with $\lambda_i^{I=C}=-q/R_i^{\;2}$), which is consistent with the fact that $\bar b_i^{T,I}$ is non-dynamic in those two sectors.

First let's review how things worked in the $N=1$ case.


\subsubsection{The $\bm{N=1}$ Coupled Diagonal Scalar Sector} \label{coupleddiagonalscalarsN1}
The $N=1$ case was studied in \cite{DeWolfe:2001nz,Bousso:2002fi} and restudied in \cite{Hinterbichler:2013kwa}.  This case is simple: there are no $\theta_{ij}$ terms, just a system of two coupled fields $\bar b^I$ and $\bar h^I=h^I-2\lambda b^I/c$ associated with a single eigenvalue $\lambda^I$.  The equations of motion for these two fields, can be written as
\begin{gather}
\label{N1matrix}
\hspace{-.3in}\Box_x\left(\begin{array}{c} \bar h\\c\bar b\end{array}\right)=
\left[\frac1{R^2} \left(\begin{array}{cc}-q\frac{q-1}{D-2} c^2R^2&0\\  0 &0 \end{array}\right) + A\right]\left(\begin{array}{c} \bar h\\c\bar b\end{array}\right), 
\end{gather}
where we've defined the $2\times2$ matrix $A$ by
\begin{gather}
\label{N1Amatrix}
A=\frac1{R^2} \left(\begin{array}{cc}  -R^2\lambda^I -2(q-1)+ q c^2R^2-2\frac{q-1}qR^2\lambda^I & -\frac{4}{c^2R^2}\frac{q-1}qR^2\lambda^I\left(R^2\lambda^I+q\right)\\ \frac{q-1}qc^2R^2 & -R^2\lambda^I + 2\frac{q-1}{q}R^2\lambda^I \end{array}\right). 
\end{gather}
(This---admittedly bizarre---way of writing it will be helpful when we move to $N>1$.)
When $\lambda^I=0$ or when $\lambda^I=-q/R^2$, only $\bar h$ is dynamic and it decouples from $\bar b$; we can write  the equation of motion out explicitly in those two cases as:
\begin{gather}
\label{zeromodeequationagain}
\Box_x  \bar h^{I=0}=\frac1{R^2}\left(-2(q-1)+ \frac{q(p-1)}{D-2} c^2R^2\right)\bar h^{I=0}, \\
\Box_x  \bar h^{I=C}=\frac1{R^2}\left(q + \frac{q(p-1)}{D-2}c^2R^2\right) \bar h^{I=C}.
\label{ellis1Nis1}
\end{gather}
Equation \eqref{zeromodeequationagain} is the zero-mode equation studied in the previous subsection.  Zero-mode stability corresponds to
\begin{gather}
\label{0modestability}
\ell=0 \text{ stability:} \hspace{.3in} c^2R^2> 2\frac{q-1}{q}\frac{D-2}{p-1}.
\end{gather}
Equation \eqref{ellis1Nis1} is the formula for the single physical mode in the conformal-scalar sector (it's the $\ell=1$ mode where flux sloshes to one side of the sphere); this mode has positive mass for all $N=1$ compactifications.  To find the masses of the higher modes, we need to compute the eigenvalues of the $2\times2$ matrix in Eq.~\eqref{N1matrix}.  

Let's first consider the case where $M_p$ is de Sitter.  In this case, stability means that all the fluctuations need to have $m^2>0$.  The condition that the eigenvalues of the $2\times2$ matrix are both positive is
\begin{gather}
\text{higher }\ell\text{, } m^2>0: \hspace{.3in} c^2R^2< \frac{\ell(\ell+q-1)-2q+2}{q-2}\;\frac{D-2}{p-1},
\end{gather}
where we use $\lambda^I=-\ell(\ell+q-1)/R^2$.  
For $q=2$, this inequality is automatic
, so all higher-mode fluctuations have a positive mass when $q=2$.  For larger $q$, this is an increasing function of $\ell$; this means that as you raise $c$ from the Nariai solution (with $c=0$), the first excited mode to develop a negative mass has $\ell=2$, then $\ell=3$, and so on.  It also means that the worst-case mode for shape-stability is $\ell=2$, for which
\begin{gather}
\label{ell2stable}
\ell=2\text{, } m^2>0: \hspace{.3in} c^2R^2< \frac{4}{q-2}\frac{D-2}{p-1}.
\end{gather}
To determine the stability of a de Sitter vacuum, there are two relevant inequalities.  First, $cR$ must satisfy Eq.~\eqref{0modestability} to evade the total-volume zero-mode instability; second, $cR$ must satisfy Eq.~\eqref{ell2stable} to evade the $\ell=2$ instability.  (In order to be de Sitter at all $cR$, must satisfy Eq.~\eqref{dSvacuum}.) Taking $p\ge3$, we find: for $q=2$ or $q=3$, de Sitter vacua are only ever unstable to the $\ell=0$ mode; for $q=4$, solutions with $cR$ near the Minkowski value have an $\ell=2$ instability, and solutions with small $cR$ have an $\ell=0$ instability, but there is an island of stability in between; for $q\ge5$, that island is engulfed and all de Sitter solutions are unstable either to $\ell=0$ or $\ell=2$ fluctuations.  These results are summarized in Fig.~2.

Next, let's consider the case where $M_p$ is AdS.  In this case, stability means that all   the fluctuations have mass squareds that are no more negative than the BF bound \cite{Breitenlohner:1982bm}: 
\begin{gather}
m^2>m^2_\text{BF}\equiv\frac14\frac{(p-1)^2}{L^2}=\frac{p-1}8\frac1{R^2}\left(2(q-1)-c^2R^2\right).
\end{gather}
The case with $\Lambda=0$ (or for any value of $\Lambda$ with $c\rightarrow\infty$) has simple eigenvalues: $cR$ is given by
\begin{gather}
\label{cRnothing}
c^2R^2\Bigl|_{\Lambda=0}=\frac{2(D-2)(q-1)}{p-1},
\end{gather}
and the eigenvalues of the $2\times2$ matrix in Eq.~\eqref{N1matrix} are
\begin{gather}
\label{eigsN1}
m^2=\frac{\ell(\ell-q+1)}{R^2} \hspace{.3in} \text{and} \hspace{.3in}  m^2=\frac{(\ell+q-1)(\ell+2q-2)}{R^2}.
\end{gather}
The second eigenvalue is always positive for all $\ell\ge2$; the first eigenvalue is negative whenever $\ell<q-1$, and is most negative when $\ell=(q-1)/2$.  We need to compare these negative mass squareds to the BF bound which, for this value of $cR$, corresponds to
\begin{gather}
\label{BFN1}
m^2_\text{BF}=-\frac{(q-1)^2}{4}\frac{1}{R^2}.
\end{gather}
When the first eigenvalue is at its most negative, $\ell=(q-1)/2$, it exactly saturates the BF bound; all other fluctuations are above the bound.  This critical value of $\ell$ is only present in the spectrum when $q$ is odd and bigger than 4.  (If $q$ is even, the critical value of $\ell$ is not in the spectrum, and if $q\le3$ it corresponds to $\ell\le1$, which is where the modes require special treatment because of residual gauge invariance.)  In summary, when $\Lambda=0$, all fluctuation modes are stable.  When $q$ is odd and bigger than 4, there is a mode that lies exactly at the BF bound, but otherwise all modes are above the bound.  

\newpage

To complete our survey of the $N=1$ case, we need to study the case when $\Lambda\ne0$.   The $c\rightarrow\infty$ `nothing state' has the same value of $cR$ and  the same eigenvalues as in the $\Lambda=0$ case, so let's start there and consider lowering $c$.
 When $\Lambda<0$, lowering $c$ only makes shape modes more stable, so all $\Lambda\le0$ compactifications are stable.  When $\Lambda>0$, however, lowering $c$ makes shape modes less stable and those modes that were previously near the BF bound can get pushed under, into the unstable regime.  For odd $q>4$, there was a mode exactly at the BF bound when $c$ was infinite and it \emph{immediately} goes unstable as you lower $c$; instability persists for all values of $c$.  For even $q>3$, there was no mode at the BF bound, so as you lower $c$ from infinity there is a window of stability before any mode goes unstable: these stable vacua are deep AdS minima.  
 When $q=3$, some modes have a negative mass squared, but it is always above the BF bound and the vacua are always stable; when $q=2$, all fluctuation modes always have a positive mass.  

In summary:
\begin{itemize}
\item The zero-mode is stable for all AdS solutions, as well as for a range of dS solutions.
\item When $q=2$, all higher-mode fluctuations have positive mass, for any $\Lambda$.
\item When $q=3$, all higher-mode fluctuations have a stable mass squared, for any $\Lambda$.
\item When $\Lambda\le0$, even for $q\ge3$, all higher-mode fluctuations have a stable mass squared.
\item When $\Lambda>0$ and $q\ge4$, most vacua are unstable. 
\end{itemize}
For more details, see Fig.~2.

\subsubsection{The $\bm{N\ge2}$ Coupled Diagonal Scalar Sector} \label{TGNCDSS}
We will show that for $N\ge2$, like for $N=1$, all shape modes of $q=2$ and $q=3$ compactifications are stable but that, unlike for $N=1$,  instabilities can appear for $q\ge4$ even when $\Lambda\le0$.  

The fluctuation equations of motion, Eqs.~\eqref{cscalar1}-\eqref{cscalar3}, can be written as 
\begin{gather}
\Box_x\left(\begin{array}{c} \bar h_1^I \\ c_1\bar b_1^{T,I}\\ \vdots\\ \bar h_N^I \\ c_N \bar b_N^{T,I}\\ \hline\bar\theta_{ij}^{LL,I}\end{array}\right) = 
\left(\begin{array}{c c c c c | c} \;&\;&\;&\;& \;&\;\\\;&\;&\;&\;& \;&\;\\\;&\;&S&\;& \;&0\\ \;&\;&\;&\;& \;&\;\\\;&\;&\;&\;& \;&\;\\\hline\; &\;&K&\;&\; &(\lambda_\text{tot} -2\frac{p-1}{L^2} )\mathds{I}\end{array}\right) \left(\begin{array}{c} \bar h_1^I\\c_1\bar b_1^{T,I}\\\vdots\\\bar h_N^I\\c_N\bar b_N^{T,I}\\ \hline \bar\theta_{ij}^{LL,I}\end{array}\right), \label{lotsofampersands}
\end{gather}
where $S$ is an $2N\times 2N$ matrix that reproduces the couplings in Eqs.~\eqref{cscalar1} and \eqref{cscalar2}; $K$ is an $N(N-1)/2\times 2N$ matrix that reproduces the couplings of $\bar\theta_{ij}^{LL,I}$ to $\bar h_i^I$, $\bar h_j^I$, $\bar b_i^{T,I}$ and $\bar b_j^{T,I}$ in Eq.~\eqref{cscalar3};  $\mathds{I}$ is the identity matrix; and $\lambda_\text{tot}=\sum \lambda_k$.

 The first thing to notice is that this matrix is block-lower-triangular: the eigenvalues of the whole matrix are the same as the eigenvalues of $S$, plus the eigenvalue $-\lambda_\text{tot} - 2(p-1)/L^2$ occurring with multiplicity $N(N-1)/2$.  These extra eigenvalues correspond to fluctuations of $\theta_{ij}^{LL,I}$ while $\bar h_i^I$ and $\bar b_i^{T,I}$ are 0; they therefore only contribute to the spectrum if $Y_i^{I_i}$ and $Y_i^{I_j}$ are excited, so that $\theta_{ij}^{LL,I}$ corresponds to a physical fluctuation of the metric.  This implies that $\lambda_\text{tot}\le-q/R_i^{\;2} -q/R_j^{\;2}$, which is more than enough to ensure that $-\lambda_\text{tot} - 2(p-1)/L^2>0$.  So, all fluctuations of $\bar\theta_{ij}^{LL,I}$ at fixed $\bar h_i^I=\bar b_i^{T,I}=0$ have a positive mass.  In our search for instabilities, we can focus on the eigenvalues of $S$.
 
 The matrix $S$ is only $2N\times2N$ if all of the $\lambda_i^I$ are excited.  If $\lambda_i^I=0$ or if $\lambda_i^I=-q/R_i^{\;2}$, then $\bar b_i^{T,I}$ is not dynamic, $\bar h_i^I$ decouples from it, and the matrix $S$ seals up by one row and one column.  Notice that taking a zero mode and promoting it to a conformal-scalar mode preserves the dimension of $S$ and only adds positive numbers down the diagonal, augmenting stability.  In other words, \emph{a mode with $\ell_i=1$ can only be unstable if the same mode except with $\ell_i=0$ is even more unstable.}

 Scalar modes can be divided into two types: modes where all of the $\ell_i$ are either 0 or 1, and shape modes where at least one of the $\ell_i\ge2$.  If a solution is stable to zero modes, it is necessarily stable to all modes of the first type.  In this section, we will investigate stability against the second type of modes.  We will prove that all shape-mode fluctuations of $q=2$ and $q=3$ compactifications are stable.  Our proof strategy will be as in Sec.~\ref{0modesN}.  First, we will identify a friendly solution and demonstrate that for it all shape-mode fluctuations are stable.  Then, we will consider paths through $c_i$ space that preserve this stability.

\paragraph{Friendly Solution: $\bm{c_k=0 \; \forall\;k}$}
\label{Nariai}
For our friendly solution, we are allowed to choose any point we like, so let's make things as easy as we can and choose the solution with $c_k=0$ for all $k$; when $\Lambda>0$, this is the Nariai solution and when $\Lambda\le0$, this corresponds to the solution where the internal $M_{q,k}$ have all blown up to infinite size and $L^{-2}=0$.  
We've already seen that the Nariai solution has an unstable zero-mode, but this will not matter for our purposes.
While these solutions are not necessarily stable to a mode where all of the $\ell_i$ are 0 or 1, we will now show that they are always stable against higher-mode fluctuations where at least one of the $\ell_i\ge2$.  For our purposes of investigating stability against these higher-modes, the Nariai solution therefore can function as our friendly anchor solution.

When $c_k=0$, the $M_{q,k}$ decouple from one another, because it was only the background flux density that was coupling the sub-manifolds.  The matrix $S$ breaks apart into $2\times2$ diagonal blocks,  and the eigenvalues of the $i$th block are
\begin{gather}
\ell_i\ge2\text{, }c_k=0\;\forall\;k: \hspace{.3in} m^2=-\lambda_\text{tot}^I, \hspace{.3in}\text{and}\hspace{.3in} m^2=-\lambda_\text{tot}^I-2\frac{q-1}{R_i^{\;2}},
\end{gather}
where the first eigenvalue corresponds to perturbations of the flux and the second corresponds to perturbations of the shape; both are positive because $\lambda_\text{tot}^I \leq \lambda_i^I \leq -2(q+1)/R_i^{\;2}$, where we've used $\ell_i\ge2$.  What about when $\ell_i=0$ or $1$?  In those cases, there is only one physical mode for that sub-manifold, and its eigenvalue is
\begin{eqnarray}
\ell=0\text{, }c_k=0\;\forall\;k: &\hspace{.3in} & m^2=-\lambda_\text{tot}^I-2\frac{q-1}{R_i^{\;2}} \\
\ell=1\text{, }c_k=0\;\forall\;k: &\hspace{.3in} & m^2=-\lambda_\text{tot}^I.
\end{eqnarray}
All of these modes are stable unless $\lambda_\text{tot}^I>-2(q-1)/R_i^{\;2}$, which cannot happen when \emph{any} of the $\ell_k\ge2$.  

\paragraph{Paths with Positive Determinant} 
 As we did in Sec.~\ref{pathszeromode}, we will use the matrix determinant lemma. 
 If all the $\lambda_i^I$ are excited (which for spheres means $\ell_i\ge2$), then the matrix $S$ is $2N\times2N$ and can be written as $A+UV^T$, where $U$ and $V$ are $2N\times1$ column vectors.  We define the $2\times2$ sub-matrix $A_i$ as
\begin{gather}
A_i=\frac1{R_i^{\;2}} \left(\begin{array}{cc} -R_i^{\;2}\lambda_\text{tot} -2(q-1) + q c_i^{\;2}R_i^{\;2}-2\frac{q-1}qR_i^{\;2}\lambda_i & -\frac{4}{c_i^{\;2}R_i^{\;2}}\frac{q-1}qR_i^{\;2}\lambda_i(R_i^{\;2}\lambda_i+q) \\ \frac{q-1}qc_i^{\;2}R_i^{\;2} &-R_i^{\;2}\lambda_\text{tot}+ 2\frac{q-1}{q}R_i^{\;2}\lambda_i  \end{array} \right),
 \end{gather}
which is analogous to the matrix $A$ defined in the $N=1$ case in Eq.~\eqref{N1Amatrix}, so that
 \begin{gather}
A=\left(\begin{array}{ccc} A_1&0&0\\0&\ddots&0\\0&0&A_N\end{array}\right), \\
 U^T= \left(\begin{array}{ccccccc} 1&0& 1&0&\cdots&1&0\end{array}\right),\\
 V^T=-q\frac{q-1}{D-2}\left(\begin{array}{ccccccc} c_1^{\;2}&0& c_2^{\;2}&0&\cdots&c_N^{\;2}&0\end{array}\right).
 \end{gather}
The determinant of S is given by $\det(A) ( 1 + V^{T} A^{-1} U)$, so
\begin{gather}
\label{uggo}
\det \; S=\sum_{i=1}^N \left\{\left(\prod_{k=1,k\ne i}^N \det\; A_k\right) \left[\frac{\det\; A_i}{N}-q\frac{q-1}{D-2}c_i^{\;2} \left(-\lambda_\text{tot} +2\frac{q-1}{q}\lambda_i\right)\right] \right\},
\end{gather}
where the sub-determinants are
\begin{gather}
\label{subdet}
\det \; A_i= \frac{c_i^{\;2}R_i^{\;2}\left(2(q-1)\lambda_i-q\lambda_\text{tot}\right)+\lambda_\text{tot}\left(2(q-1)+R_i^{\;2}\lambda_\text{tot}\right)}{R_i^{\;2}},
\end{gather}
and the term in square brackets is
\begin{gather}
\label{subdetminus}
\left[\frac{1}{NR_i^{\;2}}\left(c_i^{\;2}R_i^{\;2}\frac{p+N-2}{D-2}\left(2(q-1)\lambda_i-q\lambda_\text{tot}\right)+\lambda_\text{tot}\left(2(q-1)+R_i^{\;2}\lambda_\text{tot}\right)\right)\right].
\end{gather}

If $\lambda_i=0$, then $\bar b_i^{T,I}$ is non-dynamic, $A_i$ loses a column and a row to become a $1\times1$ matrix.  The sub-determinant becomes
\begin{gather}
\label{subdetI0}
(\det \; A_i)^{I=0}= \frac{qc_i^{\;2}R_i^{\;2}-2(q-1)-R_i^{\;2}\lambda_\text{tot}}{R_i^{\;2}},
\end{gather}
and the term in square brackets in Eq.~\eqref{uggo} gets replaced by 
\begin{gather}
\label{subdetminusI0}
\left[\frac1{NR_i^{\;2}}\left(q c_i^{\;2}R_i^{\;2} \frac{(p+N-2)}{D-2} -2(q-1) - R_i^{\;2}\lambda_\text{tot}  \right)\right]^{I=0}.
\end{gather}

A sufficient condition for $\det(S)>0$ is for all the sub-determinants and for all the terms in square brackets to be positive for all $i$, so that every term in the sum in Eq.~\eqref{uggo} is positive.   We will find conditions on the $c_i R_i$ that ensure this condition is met; we will treat the case where $\lambda_i$ is excited ($\lambda_\text{tot}\le\lambda_i\le-2(q+1)/R_i^{\;2}$) and where $\lambda_i$ is in its zero mode ($\lambda_i=0$, $\lambda_\text{tot}<0$) separately.

If $\lambda_i$ is excited, then the quantities we want to be positive are those in Eq.~\eqref{subdet} and Eq.~\eqref{subdetminus}.  If $2(q-1)\lambda_i>q\lambda_\text{tot}$ these terms are automatically positive. 
If $2(q-1)\lambda_i<q\lambda_\text{tot}$, these terms are only positive for a range of $c_i R_i$, and the tightest constraint on $c_i R_i$ comes from  Eq.~\eqref{subdet} when $\ell_i=2$ and all the other $\ell_k=0$; positivity of both terms is guaranteed by
\begin{gather}
\label{cond1}
\ell=(2,0,\dots,0), \; m^2>0: \hspace{.3in} c_i^{\;2}R_i^{\;2}<\frac4{q-2}.
\end{gather}

If $\lambda_i=0$, then the quantities we want to be positive are those in Eq.~\eqref{subdetI0} and Eq.~\eqref{subdetminusI0}.  The tightest constraint on $c_i R_i$ now comes  from Eq.~\eqref{subdetI0}, and from the mode for which $\lambda_\text{tot}$ is as close to zero as possible, meaning all the $\ell_k$ are set to 0 except a single $\ell_j=2$.  In that case,  Eq.~\eqref{subdetI0} becomes 
%
\begin{gather}
\label{cond2}
\ell=(0,2,0,\dots,0), m^2>0:\hspace{.3in}(\det \; A_i)^{I=0}=\frac1{R_i^{\;2}}\left(q c_i^{\;2}R_i^{\;2}-2(q-1)+\frac{R_i^{\;2}}{R_j^{\;2}}2(q+1)\right)>0,
\end{gather}
which, using the background equation of motion $2(q-1)R_i^{\;-2}+2(q-1)R_j^{\;-2}=c_i^{\;2}+c_j^{\;2}$ can be written,  for $q=2$ or $q=3$, as a sum of positive terms.  

This analysis is analogous to Eq.~\eqref{ell2stable} from the $N=1$ case, where positivity of the $\ell=2$ mode provided the strongest bound on $cR$ for stability of the de Sitter vacua. If both the bound in Eq.~\eqref{cond1} and the bound in Eq.~\eqref{cond2} are satisfied, then $\det(S)>0$ for all shape modes.  (If either bound is violated, we learn nothing about the sign of $\det(S)$.) 
For $q=2$, both conditions are guaranteed because the bound on $cR$ is itself unbounded; all fluctuations around solutions with  $q=2$ have positive mass.  When $q=3$, the bound in Eq.~\eqref{cond1} overlaps with the de Sitter condition in Eq.~\eqref{dSvacuumN}, so $\det(S)>0$ for all de Sitter vacua; all fluctuations around de Sitter solutions with $q=3$ are stable, but for AdS solutions, some mass squareds might go negative.  For  $q\ge4$, this proof strategy reveals no information about stability.

  Next, let's look at AdS compactifications.  In order to prove stability of solutions with $q=3$, we need to prove that, though some mass squareds may be negative,  they are never more negative than the BF bound---we need to compare the mass squareds not to zero but to  $m^2_\text{BF}$.  
  This can be accomplished by taking $S \rightarrow S - m^2_\text{BF} \mathds{I}$ and rerunning the analysis above. Following these steps proves that all fluctuations when $q=3$ are stable.



\paragraph{Instabilities for $q \geq 4$}
%
We have just shown that Freund-Rubin compactifications built of products of 2- or 3-dimensional Einstein manifolds are always stable to shape modes. The same is not true for $q \geq 4$. For a given set of $c_i$ and $\Lambda$ and for a given mode specified by a set of $\lambda^{I}_i$, stability can checked directly by evaluating the eigenvalues of the matrix $S$ defined in Eq.~\ref{lotsofampersands}. The results of this analysis are given in Fig.~3.  When $q\ge4$ the solutions for some values of $c_i$ and $\Lambda$ are stable (for example $N=2$, $p=4$, $q=5$, $\Lambda=+1$, with $c_1=c_2=10$ and so $L^{-2}=-97/9$ is stable to all fluctuations), and the solutions for other values are unstable (for example $N=2$, $p=4$, $q=5$, $\Lambda=0$, with $c_1=3$ and $c_2=4$ and so $L^{-2}=-25/18$ is unstable to the mode with $\ell_1=0$ and $\ell_2=2$).  In general, increasing $p$ aids stability (by lowering the BF bound), increasing $q$ or $\Lambda$ hurts stability (for the same reason as in $N=1$), and for a given value of $p$, $q$ and $\Lambda$ the stablest compactifications are those with all of the $c_i$ equal. Unlike in the $N=1$ case some $q=5$ and $\Lambda > 0$ minima are now stable. The greatest difference from the $N=1$ case, however, is that $\Lambda \leq 0$ no longer guarantees stability.

\subsection{The Remaining Scalar Fluctuations}
\label{offdiagonalscalars}

In this section, we will look at the remaining scalar fluctuations. First, in the diagonal $TT$ sector, there is the transverse traceless mode $\phi_i^{TT,I}$, which can be thought of as a graviton propagating on $M_{q,i}$; it is here that we'll find the `cycle-collapse instability'.  Second, in the off-diagonal $TT$ sector, there is a coupled system made up of $\theta_{ij}^{TT}$ and $\beta_{ij\cdots j}^{TT}$.  Finally, in the off-diagonal $TL$ sector, there is a coupled system made of $\theta_{ij}^{LT}$ and $\beta_{ij\cdots j}^{LT}$.
This list is complete because the rest of the scalar fluctuations are either non-dynamic, or decoupled from gravity.
Our gauge choices in Eqs.~\eqref{condition2}, \eqref{condition3}, and \eqref{condition4} essentially solve for $\phi_i^{LT}$, $\phi_i^{LL}$ and $b_i^{L}$, so they should not be thought of as dynamic variables.  (Implicit in our gauge choice is that $\beta_{ij\cdots j}^{LT}$ and $\beta_{ij\cdots j}^{LL}$ are solved for as well.)  Form fluctuations that are more than singly off-diagonal will be treated in coordinate-free notation in Sec.~\ref{higherforms}.

\textbf{Diagonal }$\bm{TT}$\textbf{ sector}: The equation of motion for $\phi_i^{TT,I}$,  Eq.~\eqref{E31}, can be written
\begin{equation}
\Box_x \phi_i^{TT,I}= \left(-\sum_{k\ne i}\lambda_k -\tau_i - 2\frac{q-1}{R_i^{\;2}}\right)\phi_i^{TT,I}.
\end{equation}
The worst-case scenario for stability is when the $\lambda_k=0$ for all $k\ne i$, so let's consider that case.  When the internal manifold is simply connected, $\tau_i^{I_i}$ satisfies a Lichnerowicz bound that $\tau_i^{I_i}\le-4(q-1)/R_1^{\;2}$, which more than ensures stability.  However, when $M_{q,i}$ is itself a product, this bound can be violated.  For instance, if $M_{q,i}=S_{q-n}\times S_{n}$, then the mode in which the $S_{q-n}$ grows, and the $S_{n}$ shrinks in a volume-preserving way, has $\tau_i=0$; for this mode, $\phi_i^{TT,I}$ has a negative mass.  For de Sitter or Minkowski compactifications, therefore, this mode is always unstable; for AdS compactifications, stability can rescued only if $\phi_i^{TT}$'s negative mass squared is less negative than the BF bound
\begin{gather}
-2\frac{q-1}{R_i^{\;2}}\ge m_{BF}^2=\frac14\frac{(p-1)^2}{L^2}.
\end{gather}
(Remember that for AdS compactifications $L^2<0$.)
In the $N=1$ case, this condition is equivalent to $q\ge9$, as discussed in \cite{DeWolfe:2001nz}; for $N\ge2$ there is some wiggle-room because the other $c_k$'s can be used to push the compactification deep into AdS, making the BF bound arbitrarily easy to be satisfied.

We should think of the `cycle-collapse instability' we found here as a residual version of the instability in the $N=1$ case.  For instance, if you wrap an 8-form around $S_2\times S_2\times S_2 \times S_2$, you get 6 fields with a negative mass squared  (four choose two `cycle-collapse instabilities').  If instead you wrap a 4-form around the first two $S_2$'s and the last two $S_2$'s separately, then you only get two fields with a negative mass---one `cycle-collapse instability' for each individually wrapped $S_2\times S_2$.  


\textbf{Off-Diagonal }$\bm{TT}$\textbf{ sector}:  Next, let's discuss the coupled system made up of $\theta_{ij}^{TT,I}$ and $\beta_{ij\cdots j}^{TT,I}$.  Eqs.~\eqref{E41} and \eqref{M31} become:
\begin{gather}
\Box_x \left( \begin{array}{c} c_j \beta_{ij\cdots j}^{TT,I} \\ \theta_{ij}^{TT,I} \\ c_i \beta_{ji\cdots i}^{TT,I}\end{array}\right)=\left[-\lambda_\text{tot} \left(\begin{array}{ccc} 1&0&0\\0&1&0\\0&0&1\end{array}\right)  + \left(\begin{array}{ccc}0&c_j^{\;2}&0\\-\kappa_j&-2\frac{p-1}{L^2}&-\kappa_i\\0&c_i^{\;2}&0\end{array}\right)\right] \left( \begin{array}{c} c_j \beta_{ij\cdots j}^{TT,I} \\ \theta_{ij}^{TT,I} \\ c_i \beta_{ji\cdots i}^{TT,I}\end{array}\right),
\end{gather}
where we've defined $\lambda_\text{tot}$ as the eigenvalue $\triangle_y Y_{\alpha_i\beta_j}^I=\lambda_\text{tot} Y_{\alpha_i\beta_j}^I$
\begin{gather}
\lambda_\text{tot}=\sum_{k=1,k\ne i,j}^N\lambda_k+\kappa_i+\kappa_j\le-2\frac{q-1}{R_i^{\;2}} -2\frac{q-1}{R_j^{\;2}},
\end{gather}
and we've used the symmetry of the $TT$ sector under exchange of $i$ and $j$.  The eigenvalues of this $3\times3$ matrix are
\begin{gather}
\label{TTanglemasses}
m^2=-\lambda_\text{tot},\hspace{.1in}\text{and}\hspace{.1in}-\lambda_\text{tot}-\frac{p-1}{L^2}\pm\sqrt{\frac{(p-1)^2}{L^4}-c_i^{\;2}\kappa_i-c_j^{\;2}\kappa_j}.
\end{gather}
Worst-case scenario for stability is for all the $\lambda_k$ with $k\ne i,j$ to be set to zero, so let's concentrate on that case.  Extremizing the negative branch of masses over $\kappa_i$ and $\kappa_j$ subject to the constraints imposed by the Lichnerowicz bound pushes the $\kappa$'s up against those constraints.  The least positive mass in this sector has $\kappa_i=-2(q-1)/R_i^{\;2}$ and $\kappa_j=-2(q-1)/R_j^{\;2}$, and this mass is still explicitly positive.  The off-diagonal $TT$ sector, therefore, contributes $3\times N(N-1)/2$ towers of massive scalars to the spectrum.

This sector has an extra zero mode when $q=1$, a structure modulus that corresponds to the angle between the sides of a flat torus.   Our results do in fact extend to $q=1$: take all the $R_i\rightarrow\infty$ because an $S_1$ has no intrinsic curvature and consider the $1$-form flux as the gradient of an axion with non-trivial winding around each cycle.  This `flux' makes the cycles want to grow so, to have a minimum of the effective potential, we must take $\Lambda<0$; this gives an AdS vacuum that is stable against total-volume fluctuations.  The Lichnerowicz bound on $\kappa$ in this case is undefined as written, but there is a vector harmonic with $\kappa=0$: a constant vector pointing uniformly along the $S_1$.  Indeed, plugging $\kappa_i=\kappa_j=0$ and the rest of the $\lambda_k=0$ into Eq.~\eqref{TTanglemasses} reveals a massless fluctuation.  When $q=1$ this sector contributes an additional massless modulus field, but for all $q\ge2$, these angles all have positive mass. 

\textbf{Off-Diagonal }$\bm{TL}$\textbf{ sector}: Finally, let's discuss the coupled system made up of $\theta_{ij}^{LT,I}$ and $\beta_{ij\cdots j}^{LT,I}$.  Eqs.~\eqref{E42} and \eqref{M32} are the relevant ones, and they contain useful information in both their longitudinal and transverse components.  For the moment, we are interested in the transverse information; the longitudinal information will be useful in Sec.~\ref{oneforms}.  To extract this information, we define:
\begin{gather}
\bar\theta_{ij}^{LT,I}=\theta_{ij}^{LT,I}+\left(\frac12\kappa_j+\frac{q-1}{R_j^{\;2}}\right)\left(\sum_{k=1,k\neq j}^N\lambda_k\right)^{-1}\phi_j^{TL,I} \\
\bar\beta_{ijj}^{LT,I}=\beta_{ijj}^{LT,I}+\frac{\kappa_j}{q-1}\left(\sum_{k=1,k\neq j}^N\lambda_k\right)^{-1}b_j^{TL,I}.
\end{gather}
These barred variables satisfy
\begin{gather}
\sum_{k=1,k\ne i}^N \lambda_i \bar\theta_{ij}^{LT,I}=0, \hspace{.3 in}\text{and}\hspace{.3in} \sum_{k=1,k\ne i}^N\lambda_i\bar\beta_{ijj}^{LT,I}=0.
\end{gather} 

\newpage

We can extract the longitudinal information from Eqs.~\eqref{E42} and \eqref{M32} by multiplying them by $-\lambda_k$ and summing over all $k\ne j$, which gives
\begin{gather}
\label{TConstraint1}
\left[\left(\Box_x+\triangle_y+2\frac{p-1}{L^2}\right)\left(\frac12\kappa_j+\frac{q-1}{R_j^{\;2}}\right)\phi_j^{TL,I}+\left(\triangle_y -\kappa_j\right)\nabla^\mu C_{\mu,j}^{T,I}+\frac{\kappa_j\triangle_y}{q-1} c_j b_j^{L,I}\right]\nabla_{\alpha_i}Y^I_{\beta_j}=0.
\end{gather}
and
\begin{gather}
\label{TConstraint2}
\left[\left(\Box_x+\triangle_y\right)\frac{\kappa_j}{q-1}c_jb_j^{L,I}+\left(\triangle_y-\kappa_j\right)c_j\nabla^\mu b_{\mu,j}^{T,I}-\left(\frac12\kappa_j+\frac{q-1}{R_j^{\;2}}\right)c_j^{\;2}\phi_i^{TL,I}\right]\nabla_{\alpha_i}\nabla_{[\beta_j}Y_{\delta_j]}^I=0.
\end{gather}
Finally, we can subtract this off to get the transverse information.  Eqs.~\eqref{E42} and \eqref{M32} become
\begin{gather}
\Box_x\left(\begin{array}{c} c_j \bar\beta_{ijj}^{LT,I} \\ \bar\theta_{ij}^{LT,I}\end{array}\right)=\left[-\lambda_\text{tot}\left(\begin{array}{cc}1&0\\0&1\end{array}\right)+\left(\begin{array}{cc} 0&c_j^{\;2}\\\kappa_j&-2\frac{p-1}{L^2}\end{array}\right)\right]\left(\begin{array}{c} c_j \bar\beta_{ijj}^{LT,I} \\ \bar\theta_{ij}^{LT,I}\end{array}\right),
\end{gather}
where we've defined 
\begin{gather}
\lambda_\text{tot}=\sum_{k=1,k\ne j}^N \lambda_k +\kappa_j\le-2\frac{q-1}{R_j^{\;2}}.
\end{gather}
Eigenvalues of this $2\times2$ matrix are 
\begin{gather}
m^2=-\lambda_\text{tot}-\frac{p-1}{L^2}\pm\sqrt{\left(\frac{p-1}{L^2}\right)^2-c_j^{\;2}\kappa_j}.
\end{gather}
As before, the worst-case scenario from the perspective of stability is the mode where all of the $\lambda_i$ with $i\ne j$ are set to zero and where $\kappa_j$ saturates its bound, but even this mode is stable.  The off-diagonal $TL$ sector contributes $2\times N(N-1)/2$  towers of massive scalars.

\subsection{Gravi-photons and One-Forms}
\label{oneforms}
In this section, we will look at the vector fluctuations.  We will see they are all stable.  First, in the $T$ sector, there is the coupled system made up of $C_{\mu,i}^{T,I}$ and $b_{\mu,i}^{T,I}$.  Second, in the $L$ sector, there is $C_{\mu,i}^{L,I}$, which will require a field redefinition to decouple it from $h_i$ and $b_i^{T}$.  And finally, in the $H$ sector, there are the tower of one-forms associated with harmonic 2-forms $C_{\mu,i}^{H,I}$.  This list is complete because the remaining of the vector fluctuations are either non-dynamic, or decoupled from gravity. Our gauge choice implicitly solves for $b_{\mu,i}^{L}$ and form fluctuations that are more than singly off-diagonal will be treated in coordinate-free notation in Sec.~\ref{higherforms}.

$\bm{T}$\textbf{ sector}:
We first consider the coupled system made up of $C_{\mu,i}^{T,I}$ and $b_{\mu,i}^{T,I}$.  Defining
\begin{gather}
\bar b_{\mu,i}^{T,I}=b_{\mu,i}^{T,I}-\frac1{q-1}\nabla_\mu b_i^{L,I},
\end{gather}
means that we can write Eqs.~\eqref{E21} and \eqref{M21} as
\begin{gather}
\text{Max}\left(\begin{array}{c}  C_{\mu,i}^{T,I}\\ c_j  \bar b_{\mu,i}^{T,I} \end{array} \right)=  \left[ -\lambda_\text{tot}  \left( \begin{array}{cc}1&0\\0&1\end{array}\right) + \left(\begin{array}{cc}-2\frac{p-1}{L^2} & -\kappa_j \\c_j^{\;2} & 0\end{array}\right) \right]  \left(\begin{array}{c}  C_{\mu,i}^{T,I}\\ c_j  \bar b_{\mu,i}^{T,I} \end{array} \right),
\end{gather}
where we've defined
\begin{gather}
\lambda_\text{tot}=\sum_{k\ne j} \lambda_k + \kappa_j\le-2\frac{q-1}{R_j^{\;2}}.
\end{gather}
Eqs.~\eqref{TConstraint1} and \eqref{TConstraint2}, together with Eqs.~\eqref{E32} and \eqref{M12}, can be used to prove that
\begin{gather}
\nabla^\mu C_{\mu,i}^{T,I}=\nabla^\mu \bar b_{\mu,i}^{T,I} = 0.
\end{gather}

Masses of the fluctuations are given by the eigenvalues of the $2\times2$ matrix above, which are:
\begin{equation}
m^2=-\lambda_\text{tot} +\frac{p-1}{L^2}\pm\sqrt{\left(\frac{p-1}{L^2}\right)^2-c_j^{\;2}\kappa_j}.
\end{equation}
When $\kappa_j$ saturates its bound, and all the other $\lambda_k=0$, the negative branch of this expression is exactly massless, and otherwise this expression is manifestly positive.
This means that for every Killing vector of the internal manifold, we find one massless vector propagating on our compactification.  Recall that our gauge fixing left the right amount of residual gauge invariance to accommodate a massless vector in the Killing-sector.  The $T$ sector contributes two towers of massive one-forms; the base of one of the towers includes one massless vector for every Killing vector.

$\bm{L}$\textbf{ sector}:
Eq.~\eqref{E22} contains useful information in its longitudinal and transverse parts.  For now, we will be interested in the transverse information, the longitudinal information will be useful in Sec.~\ref{massivegraviton}.  To extract the longitudinal information, multiply the Eq.~\eqref{E22} by $-\lambda_i$ and sum over all $i$.  This gives
\begin{equation}
\label{Hequation}
\left(\sum_{k=1}^N\lambda_k\right)\left(\nabla^\rho H_{\rho\mu}^I-\nabla_\mu H^I+\frac{1}{p-2}\sum_{k=1}^N\nabla_\mu h_k^I\right)+\sum_{k=1}^N\lambda_k\nabla_\mu \left(\frac1q h_k^I-c_kb_k^{T,I}\right)=0.
\end{equation}
Subtracting this off from Eq.~\eqref{E22} gives
\begin{gather}
\label{CLEOM}
\left(\text{Max}+\sum_{k=1}^N\lambda_k+2\frac{p-1}{L^2}\right)C_{\mu,i}^{L,I}=\frac1q\nabla_\mu\left(h_i^I-\frac{\sum_{k=1}^N\lambda_k h_k^I}{\sum_{k=1}^N\lambda_k}\right) 
-\nabla_\mu\left(c_i b_i^{T,I}-\frac{\sum_{k=1}^N\lambda_k c_k b_k^{T,I}}{\sum_{k=1}^N\lambda_k}\right). \nonumber
\end{gather}

To bring Eq.~\eqref{CLEOM} into the appropriate form for a massive vector, we define new vector field
\begin{gather}
V_{\mu,i}^L\equiv C_{\mu,i}^L-\frac{1}{\sum_{k=1}^N\lambda_k+2\frac{p-1}{L^2}}\nabla_\mu\left[\frac1q\left(h_i^I-\frac{\sum_{k=1}^N\lambda_k h_k^I}{\sum_{k=1}^N\lambda_k}\right) - \left(c_i b_i^{T,I}-\frac{\sum_{k=1}^N\lambda_k c_k b_k^{T,I}}{\sum_{k=1}^N\lambda_k}\right)\right].
\end{gather}
($C_{\mu,i}^{L,I}$ is only non-zero if at least two of the $\lambda_k$ are turned on, so the denominator of this expression is never zero.)
In terms of this new vector field, Eqs.~\eqref{CLEOM} and \eqref{E33} become
\begin{gather}
\text{Max} \; V_{\mu,i}^L = \left(-\sum_{k=1}^N\lambda_k+2\frac{p-1}{L^2}\right)V_{\mu,i}^L, \\
\nabla^\mu V_{\mu,i}^L=0.
\end{gather}
The masses of the vector fluctuations $V_{\mu,i}^L$ are therefore
\begin{gather}
m^2=-\lambda_\text{tot} +2\frac{p-1}{L^2},
\end{gather}
where
\begin{gather}
\lambda_\text{tot}=\sum_{k=1}^N\lambda_k\le \min_{i,j} \left( -\frac{q}{R_i^{\;2}}-\frac{q}{R_j^{\;2}}\right),
\end{gather}
which is at its least positive when only two modes are excited to their $\ell=1$ modes.
Even this worst-case mode has a positive mass, so the $L$ sector gives a tower of massive vectors.

$\bm{H}$\textbf{ sector}:
Finally, for the harmonic forms $b_{\mu,i}^{H,I}$, Eq.~\eqref{M23} becomes
\begin{gather}
\text{Max} \; b_{\mu,i}^{H,I} = -\left(\sum_{k\ne i}\lambda_k\right)b_{\mu,i}^{H,I}.
\end{gather}
The transversality constraint for $b_{\mu,i}^H$ can be extracted from the longitudinal part of Eq.~\eqref{M36}; multiplying Eq.~\eqref{M36} by $-\lambda_k$ and summing over all $k$ gives
\begin{gather}
\nabla^\mu b_{\mu,i}^{H,I}=0.
\end{gather} 
When all of the $\lambda_k$ with $k\ne i$ are set to zero, we find a massless vector for every harmonic two-form on $M_{q,i}$.  When the $M_{q,k}$ are excited, we find a tower of massive vectors above it.

\subsection{Massive Gravitons}
\label{massivegraviton}
Eq.~\eqref{E11} can be written as 
\begin{equation}
\left(\Box_x + \lambda_\text{tot} + \frac 2 {L^2}\right)H_{(\mu\nu)}=\nabla_{(\mu}\nabla_{\nu)} \left<H -\frac2{p-1}\sum_{k=1}^N h_k^I + 2 c_i b_i^{T,I} - \frac2q h_i^{I}\right>,
\end{equation}
where we've used Eq.~\eqref{Hequation} to eliminate divergences of $H_{\mu\nu}$, and we've defined the notation $\left<\bullet\right>$ to mean
\begin{gather}
\left<\bullet\right>=\frac{\sum_{i=1}^N \lambda_i \bullet}{\lambda_\text{tot}}.
\end{gather}  
To manipulate this equation into the form appropriate for a massive graviton, we need to define a new symmetric tensor field:
\begin{gather}
\phi_{(\mu\nu)}=H_{(\mu\nu)} +\left(\frac{1}{-\lambda_\text{tot}+\frac{p+2}{L^2}} \right)\left(\frac{p-2}{p-1}\right) \nabla_{(\mu}\nabla_{\nu)}\left<\frac{1}q h_i^I - c_i b_i^{T,I}  + \frac{1}{p-2}\sum_{k=1}^N h_k^I \right> .
\end{gather}
In terms of this new variable, Eqs.~\eqref{E11} and \eqref{Hequation} become
\begin{gather}
\Box_x\phi_{(\mu\nu)}=\left(-\lambda_\text{tot}-\frac2{L^2}\right)\phi_{(\mu\nu)} \\
\nabla^\mu \phi_{(\mu\nu)}=0.
\end{gather}
This gives a tower of massive gravitons with masses given by $m^2=-\lambda_\text{tot}>0$.  (Recall that a massless graviton propagating on curved space has an apparent mass squared of $-2/L^2$, and physical masses need to be compared against this reference value \cite{vanNieuwenhuizen:1984iz,Hinterbichler:2011tt}).  The physical masses of this tower are always positive.  The Higuchi bound for consistent propagation of a massive graviton \cite{Higuchi:1986py} on de Sitter space requires that the physical mass of the graviton exceeds
\begin{gather}
m^2\ge \frac{p-2}{L^2}.
\end{gather}
Even the first rung on the KK ladder of massive gravitons, with
\begin{gather}
\text{lightest massive graviton:}\hspace{.5in} m^2=\min_i\left(\frac{q}{R_i^{\;2}}\right)=\min_i\left(\frac{1}{R_i^{\;2}}\left(1+\frac{c_i^{\;2}R_i^{\;2}}{2}\right)\right) + \frac{p-1}{L^2},
\end{gather}
is above this bound.  All massive gravitons are stable.

\subsection{Uncoupled Form Fluctuations}
\label{higherforms}

The equations of motion for these decoupled form fluctuations is
\begin{gather}
d^\dagger d B=\left(\triangle_x+\triangle_y+d d^\dagger_{p}\right)B=0,
\end{gather}
where we used the gauge fixing condition $d^\dagger_{Nq}B=0$.  Eqs.~\eqref{M23} and \eqref{M33}-\eqref{M36}  are of exactly this form and are included in this discussion.  Because they are decoupled from gravity, we can  use a decomposition on the whole internal manifold, instead of decomposing separately for each sub-manifold like we had to do for the coupled modes.  If the fluctuation has $k$ indices along the extended manifold $M_p$, we can decompose as 
\begin{gather}
B_k=\sum_I b_k^T Y_{Nq-k}+ b^H_k Y_{Nq-k}^H,
\end{gather}
where we've used the gauge fixing condition $d^\dagger_{Nq}$ to kill the entire longitudinal component.  The equation of motion then falls apart into 
\begin{gather}
\left[\left(\triangle_x + \triangle_y -d_p d^\dagger_p\right)b_k^T\right]Y_{Nq-k} = 0, \\
\left[d^\dagger_p b_k\right] d_{Nq} Y_{Nq-k} =0, \\
\left[\triangle_x b_k^H\right]Y_{Nq-k}^H = 0.
\end{gather}

For the co-exact components, we find
\begin{gather}
\triangle_x b_k = -\triangle_y b_k, \hspace{.3in} \text{and}\hspace{.3in} d^\dagger_p b_k=0.
\end{gather}
Because the Laplacian $\triangle_y$ is negative definite, this sector contributes a tower of massive $k$-forms.

For the harmonic components, it's far simpler.    We find
\begin{gather}
\triangle_x b_k=0,
\end{gather}
meaning that there is a massless $k$-form fluctuation for every harmonic $(Nq-k)$-form.

\section{Discussion}

Freund-Rubin compactifications on product manifolds with $N=1$ can have an instability to cycle collapse, where one of the elements of the product shrinks down to zero volume.  Moving to higher $N$ cures this instability.  While collectively wrapping a higher-form flux around the entire product ($N=1$) leads to an instability, individually wrapping a lower-form flux around each element of the product ($N>1$) does not.  We have computed the spectrum of all small fluctuations around these product  compactifications, and found the conditions for stability.

 The only threats to stability arise in the scalar sector; higher-spin fluctuations are all positive semi-definite.  Within the scalar sector, the only threats are the zero modes and the coupled diagonal shape/flux system.  The results for stability are summarized in Figs.~2 and 3.  The zero-mode sector is stable for all AdS compactifications and for a range of de Sitter compactifications.  Stability of the higher-mode fluctuations depends on $q$.  All products of 2- or 3-dimensional Einstein manifolds are always stable against higher-mode fluctuations; whereas for $q\ge4$, higher-mode instabilities can exist and, when $N\ge2$, they can exist for any $\Lambda$.




When $q$ is very large, the unstable shape modes tend to have  very large angular momentum.  For example, AdS$_4\times S_{101}$, with $c$ large so that the compactification is deep in AdS and $\Lambda>0$, is stable to all fluctuations except for $\ell=50$. How can the $\ell=50$ mode be unstable while the  $\ell$ = 0, 1, 2, $\dots$ 49 and 51,  52,  $\dots$ modes are stable? Don't drums ring  higher on higher spherical harmonics?  The culprit is the coupling between the flux and shape modes, and our percussion-sourced intuitions are correct when this coupling is turned off\footnote{Another perhaps related case where the first mode to go unstable is one with high-$\ell$ is the wrinkles that form when a balloon is depressed \cite{depressedballs}.}.  For instance, flux perturbations on a fixed gravity background are stable and become increasingly stable as you raise the angular momentum $\ell$.  Likewise, shape-mode fluctuations about the uncharged Nariai solution are stable (although the zero-mode is not) and they too become increasing stable with $\ell$, as we saw in Sec.~\ref{Nariai}.  
%
%
%
%
%
The instability arises not from the flux or metric fluctuations separately, but from their coupling to each other and to the background flux: it is the off-diagonal terms in Eq.~\ref{N1matrix} that produce the negative eigenvalues, and the corresponding  eigenvectors have support on both kinds of fluctuations.

Coupling between modes is also responsible for the fact that shape modes may go unstable for any $\Lambda$ when $N>1$ even though all shape modes are stable for $\Lambda \leq 0$ when $N=1$. Equation~\eqref{cRN} makes it clear that flux wrapped around \emph{other} sub-manifolds has the same impact on $c_i R_i$ as a \emph{negative} contribution to the cosmological constant, and should therefore, applying the $N=1$ intuition, make the mode only more stable.  And yet we found the opposite.  The explanation is again the coupling between the modes: the background flux couples the shapes modes on different sub-manifolds, and the unstable eigenvector has support on all of them. When we set $c_i=0$ in Sec.~\ref{Nariai} we turned off the background flux, which turned off the coupling and restored the $N=1$ result separately for each sub-manifold.

Flow along the unstable shape-mode direction  breaks the symmetry of the internal manifold. Spontaneous breaking of Poincar\'e symmetries arises in other systems with off-diagonal coupling, for instance the Gregory-Laflamme instability of \cite{Gregory:1993vy}, the striped-phase instability of \cite{Donos:2013gda}, and even the Jeans instability. The correlated stability conjecture \cite{Gubser:2000mm} links classical instabilities like that of the shape modes  to thermodynamic instabilities.   If an endpoint of the shape-mode instability exists, it must therefore be a compactification on a warped product of lumpy spheres and its vacuum energy must be lower than the original unstable solution.  Warped compactifications on lumpy spheres have been found for the $N=1$ case \cite{Kinoshita:2007uk}, and likely exist for larger $N$.

An up-coming paper \cite{Hinterbichler:2013kwa} re-analyzes the $N=1$ case directly in the action, varying to second order in the fluctuations.  This method has the advantage of being extendable down to $p=2$, and it would be interesting to apply it to general $N$.   


Finally, all of these compactified solutions also correspond to the near-horizon limit of extremal black $(p-2)$-branes; far from the branes, spacetime is $D$-dimensional with a curvature set by $\Lambda$.  Each of our three classes of instability therefore has an interpretation in terms of an instability of a black brane.  The `total-volume instability' corresponds to the negative mode of the Nariai black hole responsible for the nucleation of charged black branes in de Sitter space \cite{Ginsparg:1982rs}.  The `cycle-collapse instability' corresponds to an instability of black branes whose horizons have non-trivial topology.  One can think of these black branes with exotic horizons as living at the tip of a cone.  For instance, we saw that $S_2\times S_2$ compactifications wrapped by a 4-form flux have an instability; to understand that instability in terms of the near-horizon limit of a black brane, consider the cone over $S_2\times S_2$ with $\Lambda=0$, which has metric
\begin{gather}
ds^2=dr^2 + r^2 d\Omega_2^{\;2}+ r^2 d\Omega_2^{\;2}.
\end{gather}
A black $(p-2)$-brane inserted at the origin $r=0$ of this metric has horizon topology of $S_2\times S_2$.  Its near-horizon limit is a compactified AdS$_p\times S_2 \times S_2$ solution that suffers from a `cycle-collapse instability'; the black brane too, therefore, has an instability for one of the spheres to grow while the other shrinks.  Finally, the `lumpiness instability' also must have an analog in the language of extremal black branes.  When $N=1$, the `lumpiness instability' only exists for $\Lambda>0$, so the analog is an instability of charged extremal black branes in de Sitter space---their horizons sprout lumps.

\section{Acknowledgements}

Thanks to Xi Dong, Steve Gubser, Kurt Hinterbichler, Ali Masoumi, Rob Myers, Gonzalo Torroba, and Claire Zukowski.

\bibliographystyle{utphys}
\bibliography{mybib.bib}

\end{document}